\documentclass[11pt,a4paper]{article}
\usepackage{jheppub}
\usepackage{graphicx,epsfig,psfrag,bm,amssymb}
\usepackage{bm}
\usepackage[dvipsnames]{xcolor}
\usepackage{braket}
\usepackage{mathrsfs,amsfonts,color}
\usepackage{hepunits}
\usepackage{hyperref}
\usepackage{comment}
\usepackage{soul}
\usepackage[caption=false]{subfig}
\usepackage{setspace}

\definecolor{linkcolor}{rgb}{0.0, 0.28, 0.67}
\hypersetup{colorlinks=true,
linkcolor=black,
citecolor=linkcolor,
urlcolor=linkcolor,
linktocpage=true
}

\newcommand{\Evec}{\mathbf{E}}

\newcommand{\be}{\begin{eqnarray}}
\newcommand{\ee}{\end{eqnarray}}
\newcommand{\E}{\mathbf{E}}
\newcommand{\B}{\mathbf{B}}

\newcommand{\jeff}{j_\text{eff}}

\newcommand{\nl}{\nonumber \\}
\newcommand{\w}{\omega}
\newcommand{\wg}{\omega_g}

\newcommand{\U}{{\mathbf U}}
\newcommand{\A}{{\mathbf A}}

\newcommand{\jveff}{\boldsymbol{j}_\text{eff}}

\newcommand{\pd}{\partial}

\newcommand{\xv}{{\bf x}}
\newcommand{\Vcav}{V_\text{cav}}

\newcommand{\nn}{\nonumber}

\definecolor{colorRTD}{rgb}{.2,.2,.7}

\definecolor{colorRTD2}{rgb}{.2,.3,.7}

\usepackage[page,toc,titletoc,title]{appendix}

\title{
Classical (and Quantum) Heuristics for Gravitational Wave Detection
}

\author[a,b]{Raffaele~Tito~D'Agnolo,}
\affiliation[a]{Institut de Physique Th\'eorique, Universit\'e Paris Saclay, CEA, F-91191 Gif-sur-Yvette, France}
\affiliation[b]{Laboratoire de Physique de l’École Normale Supérieure, ENS, Université PSL, CNRS, Sorbonne
Université, Université Paris Cité, F-75005 Paris, France}

\author[c]{Sebastian~A.~R.~Ellis}
\affiliation[c]{D\'epartement de Physique Th\'eorique, Universit\'e de Gen\`eve, 
24 quai Ernest Ansermet, 1211 Gen\`eve 4, Switzerland}

\abstract{
We derive a lower bound on the sensitivity
of generic mechanical and electromagnetic gravitational wave detectors. We consider both classical and quantum detection schemes, although we focus on the former. Our results allow for a simple reproduction of the 
sensitivities of a variety of experiments, including optical interferometers, resonant bars, optomechanical sensors, and electromagnetic conversion experiments. In the high-frequency regime, all detection schemes we consider can be characterised by their stored electromagnetic energy and the signal transfer function, which we provide. 
We discuss why high-frequency gravitational wave searches are especially difficult and primordial gravitational wave backgrounds might not be detectable above the sensitivity window of existing interferometers.
}

\begin{document}

\maketitle

\onehalfspacing
\section{Introduction}
In this work we develop a general framework to estimate the smallest detectable strain in a generic gravitational wave detector. We apply our formalism to the detection of stochastic backgrounds of primordial gravitational waves (GWs) with a focus on high-frequency ($f > \text{kHz}$) signals. Our calculations ultimately show that 
improving on existing cosmological bounds much above the frequencies probed by current interferometers might not be feasible without substantial improvements in technology. 
A provocative
analogy can be drawn between building a GUT-scale (i.e. $10^{15}$~GeV) particle collider and building a GW detector capable of detecting a signal originating when the Universe was at a GUT-scale temperature of $T\sim 10^{15}$~GeV. 
Taking current technology to build a GUT-scale collider would require extra-planetary resources. Clearly, one therefore needs a dramatic change in technology to make such a collider feasible on Earth. Similarly, to detect the GUT-scale GW signal would require scaling up resources of classical GW detectors by an amount that appears unfeasible (see e.g. Section~\ref{sec:conclusion}). A dramatic shift in technology, such as the improved manipulation of a macroscopically large number of quantum states, is therefore required.
That said, even making extreme assumptions on our ability to harness quantum states might prove insufficient. For example, while operating a LIGO-sized interferometer with a kW laser at the Heisenberg limit~\cite{Giovannetti:2011chh, RevModPhys.90.035005} outperforms a classical LIGO-sized interferometer with a laser drawing $1\%$ of the US grid capacity, it would only be sensitive to primordial GW backgrounds at $f\lesssim \text{MHz}$. 
The same is true for other detection schemes, more suited than interferometers to the high-frequency GW limit. 
Note that we only apply our formalism to estimate our ability to detect cosmogenic GWs. There are numerous sources of late-universe high-frequency GWs~\cite{Brito:2015oca, Arvanitaki:2010sy, Arvanitaki:2014wva, Casalderrey-Solana:2022rrn, Franciolini:2022htd, Franciolini:2022ewd}, discussed at length in Ref.~\cite{Aggarwal:2020olq}, which are not subject to the same issues we discuss here. 

The detection of GWs by the LIGO collaboration~\cite{LIGOScientific:2016aoc} was the culmination of decades of effort to reach the classical detection limits in interferometers. The success of LIGO, VIRGO and Kagra (LVK) has led to a renaissance in attempts to quantify theoretically well-motivated GW signals across the full frequency spectrum, ranging from signals at nHz frequencies, detectable by pulsar timing arrays~\cite{NANOGrav:2023gor,Xu:2023wog,Reardon:2023gzh,EPTA:2023fyk}, to frequencies well above a kHz~\cite{ Ghiglieri:2015nfa,Giovannini:2019oii,Ringwald:2020ist,Ghiglieri:2022rfp,Giovannini:2023itq, Brito:2015oca,Arvanitaki:2010sy,Arvanitaki:2014wva, Casalderrey-Solana:2022rrn, Franciolini:2022htd,Franciolini:2022ewd,hooper2020hot, Arvanitaki:2012cn}. Naturally, the improved understanding of well-motivated signals has also led to the development of various new ideas for how to detect gravitational waves outside of the most sensitive regions of existing interferometers, in particular below a Hz~\cite{Janssen:2014dka, Namikawa:2019tax,CMB-S4:2020lpa,Badurina:2019hst,Abe:2021ksx,AEDGE:2019nxb,Hild:2010id, Punturo:2010zz,LIGOScientific:2016wof,LISA:2017pwj,Yagi:2011wg} and above a kHz~\cite{Goryachev:2014yra,Ejlli:2019bqj,Aggarwal:2020umq,Berlin:2021txa,Domcke:2022rgu,Berlin:2023grv, Bringmann:2023gba, Domcke:2023bat,Tobar:2023gvp,Kahn:2023mrj,Domcke:2024mfu,Carney:2024zzk, Domcke:2024eti,Capdevilla:2024cby}.

In this paper, we show that the fundamental detection limit of any detector, including LVK, can be understood in terms of its stored electromagnetic (EM) energy. Practically, this limit is operative either at high frequencies, or for a sufficiently massive detector. Perhaps counterintuitively, we show that classically the optimal sensitivity obeys the same scaling with stored energy for detectors that measure signals quadratic in the GW strain $h$ and for those sensitive to signals linear in $h$. 

Our analysis reduces the sensitivity of a detector to the efficiency with which a GW converts its stored EM energy into signal energy. This efficiency is usually a frequency-dependent quantity, and we therefore borrow from the engineering terminology and refer to it as a \emph{transfer function}. We give examples of transfer functions as computed for existing detectors.

We apply our results to quantify the prospects for detecting stochastic gravitational wave backgrounds at high frequencies. Signals at high frequencies are expected to be mostly Standard Model background-free, and to offer a unique window on the primordial Universe, potentially probing high energy phenomena that are not directly accessible in the laboratory. They are therefore a focus of intense recent interest (see, e.g.,~\cite{Aggarwal:2020olq} for a review). As a result, many approaches for detecting high-frequency gravitational waves have been proposed~\cite{Goryachev:2014yra,Ejlli:2019bqj,Aggarwal:2020umq,Berlin:2021txa,Domcke:2022rgu,Berlin:2023grv, Bringmann:2023gba, Domcke:2023bat,Tobar:2023gvp,Domcke:2024mfu,Carney:2024zzk, Domcke:2024eti}.  
We show that our heuristic arguments reproduce the estimated sensitivities of these proposals.

Ultimately, our analysis demonstrates a well-known difficulty of detecting high-frequency primordial GWs. We can summarise this difficulty in one sentence: detectors measure \emph{strain}, but detecting stochastic GWs requires measuring an \emph{energy density}.
If a detector can measure a fixed energy density in GWs at frequency $\w_0$, thanks to its stored EM energy $U_0$, it needs a much larger amount of energy $U = U_0 (\w/\w_0)^3$ to make the same measurement at $\w > \w_0$.\footnote{The scaling with $\w^3$ is valid for a signal spectrum with energy density $d\rho/d\log\w \sim {\rm const.}$, but the argument that $U$ grows with a positive power of $\w$ is valid in general.}
The result is that it is prohibitively difficult to probe primordial stochastic GWs at high frequencies. We comment on what would be required of detectors operating at the classical limit of quantum noise, or ``standard quantum limit'' (SQL), to reach interesting stochastic GW parameter space at frequencies above LVK.\footnote{The terminology ``SQL'' can be used to denote the quantum-noise limit of a detector employing purely classical resources (see, e.g.,~\cite{Giovannetti:2011chh,Escher:2011eff}). However, in certain contexts, the SQL specifically designates the point at which quantum measurement imprecision and backaction noise are both minimised~\cite{clerkIntroductionQuantumNoise2010,Aspelmeyer:2013lha,Beckey:2023shi}. When we use the terminology SQL, we mean a sensitivity that scales as $1/\sqrt{N}$, where $N$ quantifies the resources used for detection, e.g., photon number, etc.} 
Improving on cosmological bounds, namely constraints on the total energy density of radiation in the Universe at BBN or CMB epochs~\cite{Mangano:2011ar, Cyburt:2015mya, Peimbert:2016bdg, Planck:2018jri}, with a realistic setup almost certainly requires going beyond the SQL. Therefore we discuss the expected sensitivity of existing detectors at their ultimate quantum limit, as for example the Heisenberg limit of interferometers~\cite{Giovannetti:2011chh,Escher:2011eff, RevModPhys.90.035005}. Despite a significant improvement over the the SQL, given a fixed amount of stored energy, we did not find any realistic (or very optimistic) detector that can go beyond cosmological bounds~\cite{Mangano:2011ar, Cyburt:2015mya, Peimbert:2016bdg, Planck:2018jri} at frequencies much above LVK. 

Throughout the paper we make an unusual choice of units, keeping $c=1$ as in natural units, but showing explicitly powers of $\hslash$. This makes our general formulas in Section~\ref{sec:minimal} manifestly dimensionless and easy to read, immediately highlighting quantum effects, but avoids cluttering the Eqs. in Section~\ref{sec:transfer} and below.

The paper is organized as follows, in Section~\ref{sec:mech} we give a simple argument that shows why the relevant figure of merit for the ultimate sensitivity is the electromagnetic energy in the detector readout rather than the (often) much larger mechanical energy in its test masses. In Section~\ref{sec:minimal} we derive general expressions for the minimal detectable strain in any classical or quantum detector. In Section~\ref{sec:transfer} we show that a simple toy model can describe the relevant features of existing detectors and give their transfer functions. In Section~\ref{sec:signals} we review the basics of primordial cosmological backgrounds of GWs. We conclude in Section~\ref{sec:conclusion} by showing numerical results and discussing the difficulties of detecting very high-frequency GWs.

\section{Mechanical Energy versus Electromagnetic Energy}
\label{sec:mech}

\begin{figure}[t]
\centering
\includegraphics[scale=0.4]{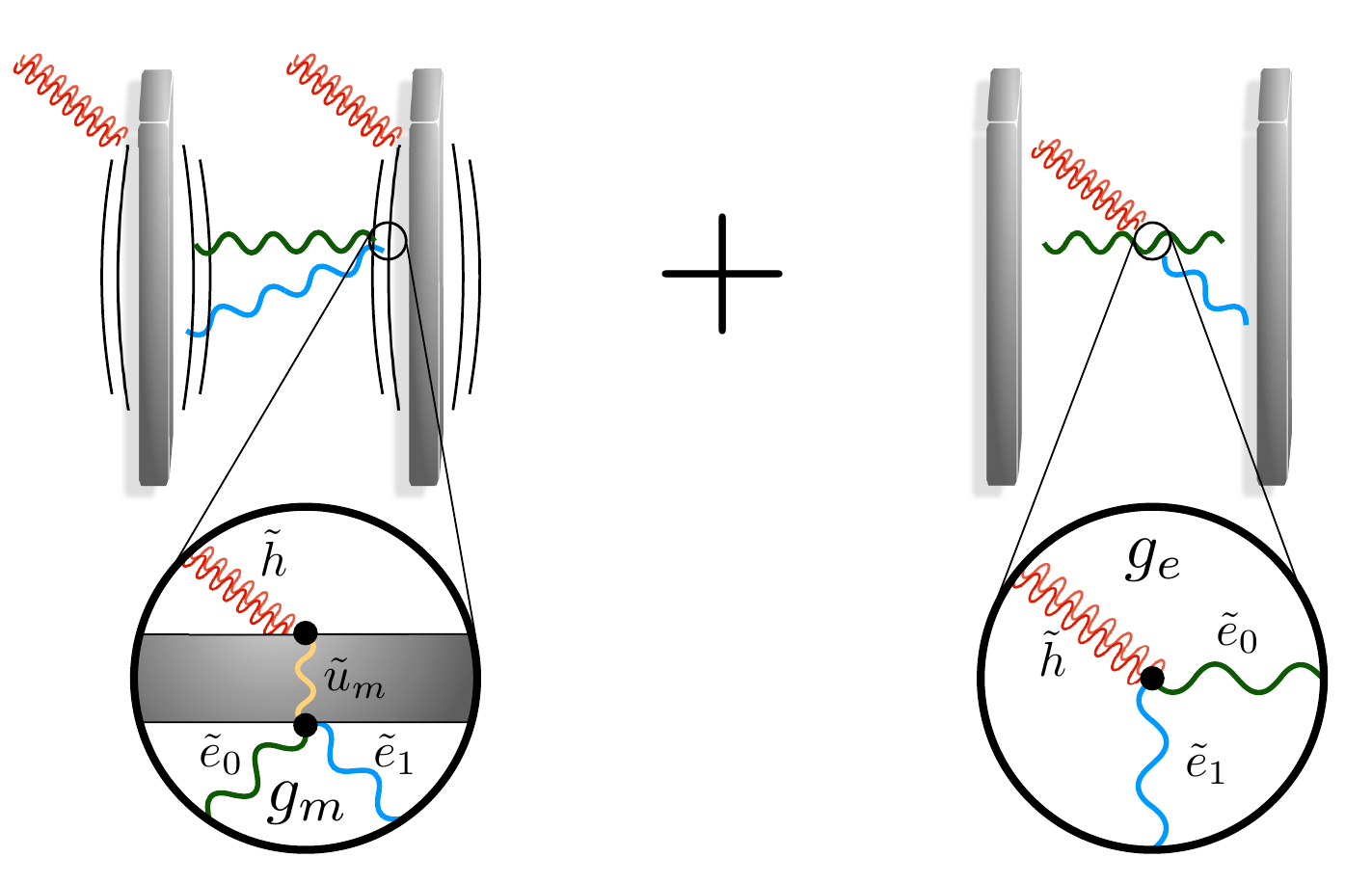}
\caption{A cartoon of the GW signals in a prototypical detector. On the left-hand side we see the familiar interaction of a GW (red wavy lines) with a test mass (grey slab). When we discuss specifics of detection in Section~\ref{sec:transfer}, this interaction corresponds to the first line of Eq.~\eqref{eq:EOM}.  On the right-hand side we see the interaction of a GW with a background EM field (green wavy line), corresponding to the second line of Eq.~\eqref{eq:EOM}. The relevant mechanical ($g_m$) and electric ($g_e$) couplings are given in Eqs.~\eqref{eq:mechCoupling} and~\eqref{eq:eCoupling} respectively. All notation is defined in Section~\ref{sec:transfer}.}
\label{fig:Cartoon}
\end{figure}

It is much easier to store a large amount of mechanical energy in a detector, by increasing its mass, than to store the same amount of EM energy. Therefore, when computing the smallest possible detectable strain $h_{\rm min}$, we typically assume we can take the mass of the detector $M$ to infinity and focus on the limitations due to the finite EM energy in the readout. In practice the most sensitive detectors now available operate in this limit. They have an EM readout of a displacement signal and they are ultimately limited by their EM stored energy. In this Section we discuss what we mean by the $M \to \infty$ limit for signal and noise. 

Our toy models for existing detection schemes are depicted in Fig.~\ref{fig:Cartoon}. As we show more explicitly in Section~\ref{sec:transfer}, we can describe the most sensitive existing detectors either as a test mass whose position is measured by an EM readout system (left panel of Fig.~\ref{fig:Cartoon}), or as a box full of EM energy (right panel).\footnote{These two toy models were also considered in~\cite{Beckey:2023shi} where they were studied in a different parametric limit.} When we take the test mass to infinity the figure of merit for GW detection in the first toy model reduces to its stored EM energy in the readout, as is already obviously the case for the second one. 

To see this, we first consider the signal produced by the GW in the detector. Mechanical energy is not important to the signal, because of the equivalence principle. A GW coupling to a test mass always leads to an equation for its displacement $\delta x$ of the form
\be
 \delta\ddot{x} \sim -\frac{1}{2}\ddot{h}\,\frac{M_{\mathsf{g}}}{M_{\mathsf{i}}} \eta L + \frac{F_{\rm mech}}{M_{\mathsf{i}}} \ , 
 \label{eq:MechEOM}
\ee
where $\ddot h$ are derivatives of the small fluctuations $h(t)$ of the metric around a flat Minkowski background (see Appendix~\ref{app:EOM} for a more detailed description of the force induced by a GW on mechanical detectors). $F_{\rm mech}$ encapsulates additional forces that could act on the detector, such as internal mechanical or external applied forces, and can produce noise in the experiment. Here $L$ roughly denotes the size of the detector. The parameter $\eta$ gives the strength of the GW coupling to the displacement, which depends on the geometry of the detector and is typically $\lesssim 1$.
The equivalence principle dictates that the inertial $({\mathsf{i}})$ and gravitational $({\mathsf{g}})$ masses are identical, $M_{\mathsf{i}} = M_{\mathsf{g}}$, such that the effect of the GW is independent of the target mass. Therefore, the signal does not benefit from having a very massive resonator. 
Instead, it is the noise that dictates that Weber bars and the LVK mirrors should be made very massive. The reason is clear from Eq.~\eqref{eq:MechEOM}: if $M_{\mathsf{i}}$ is large, it suppresses the effect of $F_{\rm mech}$ to a level that the effect of $h$ might be observable. 

If we now consider taking the limit of inertial mass $M_{\mathsf{i}} \to \infty$, the effect of $F_{\rm mech}$ goes to zero, while the GW effect does not decouple, as $M_{\mathsf{g}}\to \infty$ in the same way as the inertial mass. 
We will conduct a thought experiment where the mass can be taken to be so large that it is effectively infinite, while the volume of the detector remains fixed, without creating a black hole.
We must then consider how the resulting displacement is measured. It is most straightforward to measure this displacement by imprinting its effect on an EM field subject to boundary conditions. The boundary conditions on a region of length $L$ lead to a non-zero electro-mechanical coupling\footnote{It is easy to derive this result by noticing that the frequencies $\w_i$ of the normal modes of the EM field in a region of size $L$ depend on $L$, so $\w_i(L+\delta x) \simeq \w_i(L) +\frac{\partial \w_i}{\partial x}\delta x = \w_i(L)+g_{\rm e/m} \delta x$.} $g_{\rm e/m} \sim \w/L$, such that the classical equation of motion for the electric field, at leading order in $\delta x$, becomes
\begin{align}
    \left(\pd_t^2 + \w_i^2 + i \frac{\w \w_i}{Q_i}\right) \mathbf{E}_i(\mathbf{x},t) \simeq -2 g_{\rm e/m}^i\, \w_i \, \delta x(t) \, \mathbf{E}_0(\mathbf{x},t) \ .
    \label{eq:EoM_EM}
\end{align}
In this expression, $i$ labels the EM mode associated to the unperturbed boundary conditions and $\mathbf{E}_0$ the background EM field present in the readout in absence of a signal. With $\delta x \sim - \eta\, h\, (L/2) e^{i \w_g t}$ as the solution of Eq.~\eqref{eq:MechEOM} in the infinite mass limit for a monochromatic GW, we find that the signal EM field in frequency space is approximately
\begin{align}
    \mathbf{\widetilde E}_i(\mathbf{x},\w) \sim \eta\, h\, \frac{\w_i^2}{\w_i^2-\w^2+i \frac{\w \w_i}{Q_i}}\,\mathbf{\widetilde E}_0(\mathbf{x},\w\pm\w_g)\ . 
\end{align}
We have used that $g_{\rm e/m} \sim \w_i/L$ in obtaining this solution. It is worth emphasising at this stage that this result indicates the lack of dependence on mechanical energy of the GW-induced signal. The appearance of mechanical energy in sensitivity limits is due to noise, as discussed above.
Indeed, if we were to relax the infinite mass limit, we would introduce the contribution to $\delta x$ from $F_{\rm mech}$ in Eq.~\eqref{eq:EoM_EM}, and we would restore the expected dependence of the signal-to-noise ratio on the mechanical energy of the test mass being measured. However, it is clear that as long as the GW-dependent term in Eq.~\eqref{eq:MechEOM} dominates over the $F_{\rm mech}$ term, we are justified in ignoring the mechanical energy when estimating the sensitivity of an experiment to GWs. In what follows, we mainly focus on EM energy when quantifying the expected sensitivities of GW detectors.

Many detectors do not operate in this limit, and their sensitivity is limited by $F_{\rm mech}/M_{\mathsf{i}}$, but in this work we are interested in making the most optimistic assumptions and deriving the ultimate possible sensitivity of a detector. We are going to see in Section~\ref{sec:conclusion} that even in this often unrealistic limit most primordial sources of high frequency GWs remain out of reach.

\section{Anatomy of the Smallest Detectable Strain}\label{sec:minimal}
The smallest detectable strain of a primordial stochastic background of GWs can be computed from the Signal-to-Noise Ratio (SNR)
\be
\text{SNR}=\left(t_{\rm int} \int \frac{d\w}{2\pi} \frac{S_h^2(\w)}{S_n^2(\w)}\right)^{1/2} \ ,\label{eq:SNR}
\ee
which we derive in Appendix~\ref{app:stat} from first principles. The quantity $t_{\rm int}$ is the total integration time of the experiment and $S_{h, n}$ are Power Spectral Densities (PSDs) of the strain $h$ and the noise $n$. The general definition of a PSD can be found in Appendix~\ref{app:PSD}, the specific ones for $h$ and $n$ are discussed in Appendix~\ref{app:stat}. We do not give them here because we find more useful to give their relation to the PSDs $S_{\rm sig, noise}$ which are actually measured. When integrated over frequency $S_{\rm sig, noise}$ give the total signal and noise power, 
\be
P_{\rm sig}=\int_{-\infty}^{+\infty} \frac{d\w}{2\pi}\; S_{\rm sig}(\w),
\ee
as read out by a detector that measures EM energy. 
The SNR in Eq.~\eqref{eq:SNR} can be used as a test statistic, if compared to the appropriate quantile of its probability distribution, to establish a discovery or set an exclusion. For long enough observation times $t_{\rm int}$, we set ${\rm SNR} \simeq 1$ to estimate the signal strength (the strain in the case of a GW) that we are sensitive to in an experiment, as derived in Appendix~\ref{app:stat}.  

Other types of signals, such as black hole inspirals that can be detected using matched filtering, have different properties that allow for a more favorable scaling with $t_{\rm int}$ that we briefly discuss in Appendix~\ref{app:stat}. However in this work we are interested in primordial stochastic backgrounds, and we will discuss other signals in a companion paper~\cite{US}. A detection technique based on single photon counting~\cite{Vermeulen:2024vgl, McCuller:2022hum} allows to construct a SNR with a more favorable scaling with $t_{\rm int}$ compared to Eq.~\eqref{eq:SNR} also for a stochastic signal. We discuss this possibility below in the context of quadratic signals.

To compute the relation between $S_{h, n}$ and $S_{\rm sig, noise}$, and therefore ultimately the SNR, we can focus only on the EM energy stored in the detector, as discussed in the previous Section. The signal energy in the presence of the GW can be written, broadly, as
\begin{align}
\label{eq:Ustored}
   \langle U_{\rm sig} \rangle 
   \sim \langle E_0(t) E_h(t) \rangle V + \langle E_h(t) E_h(t) \rangle V \ ,
\end{align}
where $E_0(t)$ is the unperturbed EM field in the detector, while $E_h(t) \sim E_0(t) h(t)$ is the EM field perturbed at leading order in the strain $h$ (i.e. the dimensionless number characterizing the small fluctuations of the metric $h(t)$ around a flat Minkowski background). The angular brackets represent an average over many oscillations of the fields.
The GW can of course interact with both the EM field and the volume of the detector, so in principle there can be more terms than are given above. However volume effects are also read out by the EM field, giving signals proportional to $\langle U_{\rm sig}\rangle$, and we include them in the transfer functions introduced later in this Section and computed in Section~\ref{sec:transfer}. From Eq.~\eqref{eq:Ustored} we see that there are two qualitatively different types of signal, a linear-in-strain signal and a quadratic one. The signal targeted by the detector determines both the form of $S_{\rm sig}$ and that of $S_{\rm noise}$. We treat the two cases separately in the next Subsections.

First it is worth to comment on two aspects of our estimates. All our calculations, including Eq.~\eqref{eq:Ustored} are performed in the {\it Proper Detector Frame} (PDF), a frame of Fermi normal coordinates with origin fixed at the center of mass of the detector~\cite{Manasse:1963zz,Misner1973,Maggiore}. This is the reference frame that reproduces the experimenters' Newtonian intuition in their laboratory and makes all quantities in this paper the same as those measured in the laboratory. This choice makes the GW more complicated than in TT gauge, but all other experimental parameters much simpler. For more details we refer to~\cite{Berlin:2021txa}.

Finally, given that this simplified description was obtained under the assumption of an infinite mass detector, one might worry that our conclusions might change upon restoring the finite mass. We show below that our description is appropriate also in situations where the GW can be thought of as coupling predominantly to the mechanical energy stored in the detector, as is the case for optomechanical sensors, interferometers and Weber bars.
 
\begin{figure*}[!t]
\begin{center}
\includegraphics[width=0.45\textwidth]{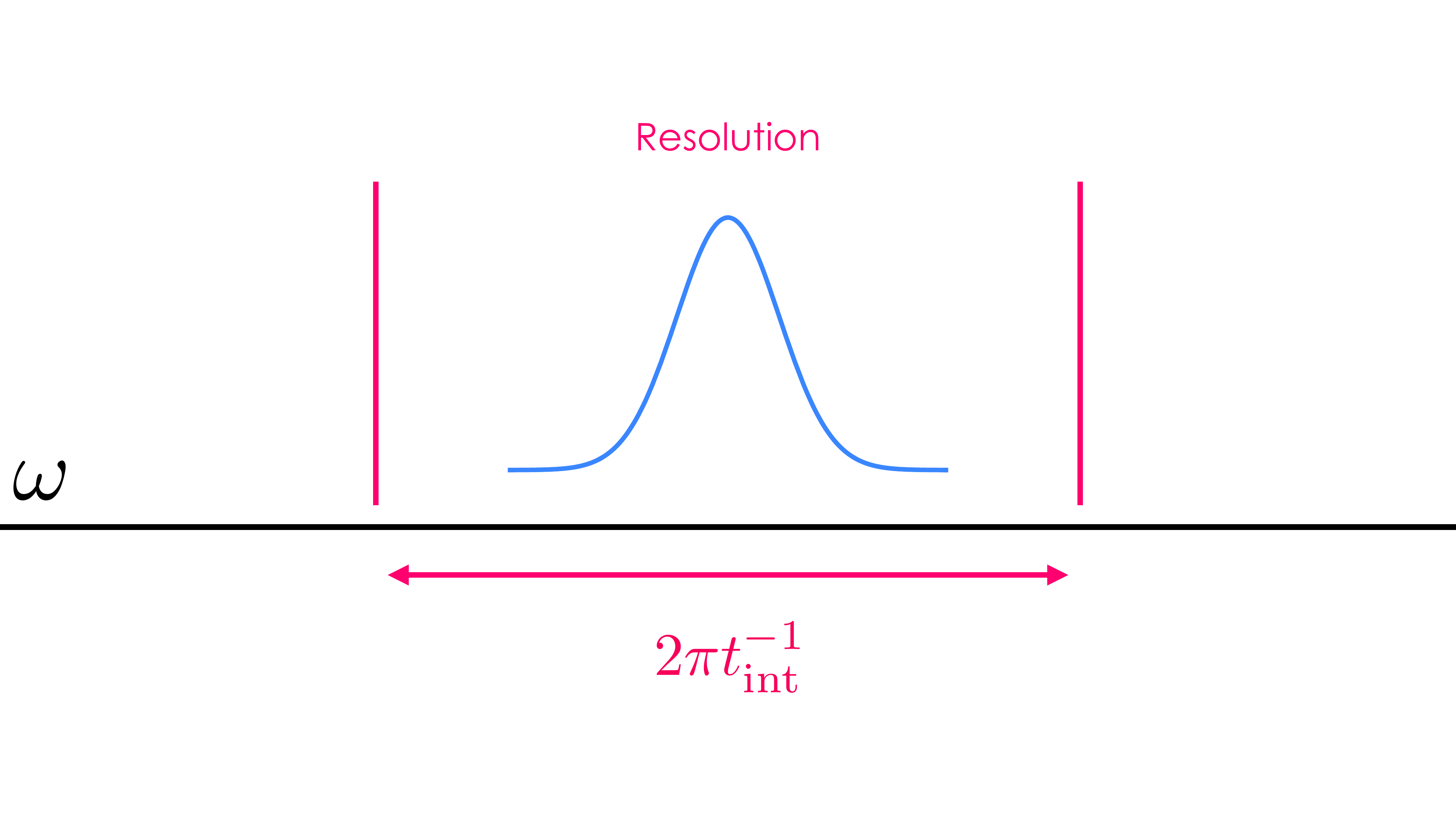}
\includegraphics[width=0.45\textwidth]{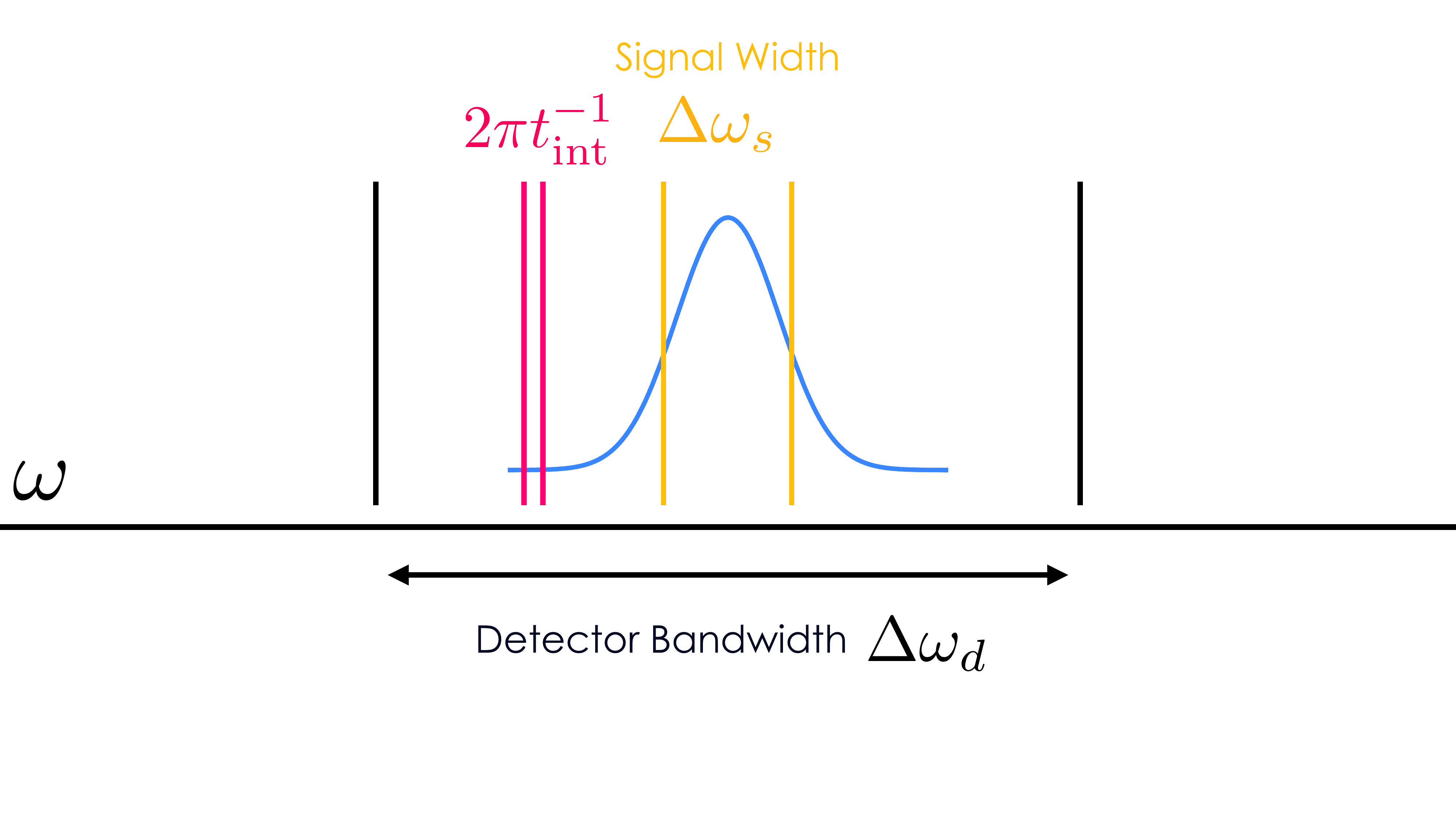}
\caption{An experiment running for a time $t_{\rm int}$ can resolve only frequencies larger than $t_{\rm int}^{-1}$. On the left $t_{\rm int}^{-1}$ is large compared to the signal width (blue curve) and detector bandwidth (outside of the plot), so
we can only measure quantities in a single bin $\Delta \nu= t_{\rm int}^{-1}$. On the right we see the opposite situatio. The detectable signal is in a range $\Delta \nu=\min[\Delta \nu_d, \Delta \nu_s] \gg 1/t_{\rm int}$ and we can resolve  $\Delta \nu t_{\rm int}$ frequency bins containing the signal.}
\label{fig:width}
\end{center}
\end{figure*}

\subsection{Quadratic Signals}
Let us first examine the quadratic term in Eq.~\eqref{eq:Ustored}. 
The quantity $\langle E_0(t) E_h(t)\rangle$ need not be zero, but if it is not, then we are clearly examining the sub-dominant signal. Therefore, let us perform a thought experiment where this term is zero. This describes a variety of existing detectors examined in Section~\ref{sec:transfer}. For this kind of signal we have the relation
\be
\frac{S_h(\w)}{S_n(\w)}=\frac{S_{\rm sig}(\w)}{S_{\rm noise}(\w)}\, ,
\ee
and the sensitivity of a detector can be estimated from 
\be
\text{SNR}=\left(t_{\rm int} \int \frac{d\w}{2\pi} \frac{S_{\rm sig}^2(\w)}{S_{\rm noise}^2(\w)}\right)^{1/2} \simeq 1\, .
\label{eq:SNRquad}
\ee
If $\langle E_0(t) E_h(t)\rangle = 0$, then $\tilde E_0(\w)$ and $\tilde E_h(\w)$ do not have support in the same frequency ranges, because 
\be
\langle E_0(t) E_h(t) \rangle &=& \lim_{T\to \infty}\frac{1}{2T}\int_{-T}^T dt\, E_0(t) E_h(t) \propto \int \frac{d\w}{2\pi} \,\langle \tilde E_0(\w) \tilde E_h(\w) \rangle \, .
\ee
Here we denote with $\tilde E$ the Fourier transform of $E$ and the angular brackets in frequency space represent an ensemble average, which is needed for the stochastic primordial signals that we are interested in~\cite{Maggiore_2007}. 
Since $\tilde E_0$ and $\tilde E_h$ do not have support in the same frequency ranges, for all frequencies where $\langle \tilde E_h^2(\w)\rangle \neq 0$ we must have $\langle \tilde E_0^2(\w) \rangle = 0$, i.e. the signal is localized away from the unperturbed energy in frequency space. 
If our detector looks in a small enough frequency window around the signal, we are left with a signal power that is quadratic in $h$, and the noise power should not depend on the injected energy, since $\langle \tilde E_0^2(\w) \rangle = 0$ in the signal region. If there is no signal, i.e. $h=0$, then the power seen by the detector should be zero, classically. 
Quantum-mechanically we have to account for zero-point fluctuations of the vacuum. The minimal noise PSD in a quadratic detector at a temperature $T$ is therefore
\be
S_{\rm noise}^{\rm quad}(\w) =\frac{\hslash\w}{1-e^{-\hslash\w/k_B T}}\, .
\label{eq:Pnoise}
\ee
In the zero-temperature limit, this amounts to a single photon. 
Now we can introduce a transfer function $\mathcal{T}^{\rm quad}$, to model the way the detector transforms the GW energy into an EM signal, 
\be
\langle \tilde U_{\rm sig}(\w)\rangle \equiv h^2 U_{\rm in}\left(\mathcal{T}^{\rm quad}(\w)\right)^2, 
\label{eq:Usig}
\ee
where we have defined the total stored EM energy in the detector, in the absence of a GW, as $U_{\rm in}\equiv \sqrt{\langle U_0^2(t)\rangle}$. In the above expression we have chosen to factor out the dimensionless number $h$ that characterizes the strength of the GW. A precise definition of $h$ can be found in Appendix~\ref{app:energy}. 
The choice that we made in Eq.~\eqref{eq:Usig} moves to the transfer function all the details of the detection scheme. This includes potentially large factors such as the enhancement $L/\lambda$ that one obtains in an interferometer from the small wavelength of the laser $\lambda$ and the large size of the arms $L$. This number is of $\mathcal{O}(10^{11})$ for LVK. The factors in $\mathcal{T}$ also include a function of $\w_g L$ coming from the Riemann tensor, that gives a different parametric scaling in the two limits where $\w_g L \ll 1$ (long wavelength limit) or $\w_g L \gg 1$. Calling these factors a detector effect or a GR effect is just a semantic choice. We chose to include them in $\mathcal{T}$ to highlight some features of $h_{\rm min}$ that we derive in this Section, and contain important physical information whose discussion is absent elsewhere in the literature.
However, one should keep in mind that extracting a valid numerical estimate of the minimal detectable strain $h_{\rm min}$ requires knowing the detector's details encoded in $\mathcal{T}$ and we postpone this to Section~\ref{sec:transfer}.

To compute the signal PSD we make a maximally optimistic assumption, we take $S_{\rm sig}$ to be constant and equal to its maximal value $S_{\rm sig}(\w_s)$ in the whole detection bandwidth of the experiment $\Delta \w$. With this assumption we can write
\be
S_{\rm sig}^{\rm quad}(\w)&\leq&h^2 U_{\rm in}\left(\mathcal{T}^{\rm quad}(\w_s)\right)^2\frac{\w_s}{\Delta \w}\left[\Theta\left(\w-\w_s+\frac{\Delta \w}{2}\right)-\Theta\left(\w-\w_s-\frac{\Delta \w}{2}\right)\right]\, . \label{eq:Psig}
\ee
Saturating the equality corresponds to the largest possible PSD compatible with Poynting's theorem (if $\Delta \w$ is small enough), as it is assuming a timescale for energy  transfer $t=2\pi/\w_s$ for a signal oscillating at $\w_s$. A laser, for instance, saturates the inequality. As we will see in Section~\ref{sec:transfer}, for many other detectors the signal PSD can be suppressed by a ratio as small as $\Delta \w/\w_s$. In the above, we define the detection bandwidth as  
\be
\Delta \w\equiv\max[\min[\Delta \w_d, \Delta \w_s],(2\pi)\,t_{\rm int}^{-1}]\, . \label{eq:domega}
\ee
It is given by the minimum of the signal width $\Delta \w_s$ and detector bandwidth $\Delta \w_d$, compared to the resolution in frequency space $(2\pi)\,t_{\rm int}^{-1}$. This choice of bandwidth is explained in Fig.~\ref{fig:width} and can be articulated as follows. The signal has support in the smallest interval between $\Delta \w_d$ and $\Delta \w_s$, because we can only  detect the fraction of the signal within the detector bandwidth $\Delta \w_d$. Then we have to compare $\min[\Delta \w_d, \Delta \w_s]$ with the smallest frequency  interval that we can resolve, which is given by $(2\pi)\,t_{\rm int}^{-1}$. If our integration time is short and $(2\pi)\,t_{\rm int}^{-1}$ is the largest frequency interval in the problem we will just see all the signal in a single bin of size $(2\pi)\,t_{\rm int}^{-1}$. In the opposite limit we are looking at a signal spread over $N_b \simeq \min[\Delta \w_d, \Delta \w_s] t_{\rm int}/2\pi$ frequency bins.

We are now ready to estimate the minimal detectable strain in a classical experiment, using Eq.s~\eqref{eq:SNRquad}, \eqref{eq:Pnoise} and \eqref{eq:Psig}
\be
    h_{\rm min,\ quad}^C &\gtrsim& \sqrt{\frac{\hslash}{U_{\rm in}}}\left(\frac{2\pi\Delta \w}{ \,t_{\rm int}}\right)^{1/4} \frac{1}{\mathcal{T}^{\rm quad}(\w_s)} \label{eq:quad}\, . 
\ee
Eq.~\eqref{eq:quad} gives the smallest detectable strain in a classical experiment, i.e. an experiment whose classical noise sources have been made negligible and operates at its SQL. 
While the single photon of noise from quantum fluctuations of the vacuum is always present, we need not necessarily measure it. The noise power in a measurement employing quantum resources can be reduced, at least conceptually, to almost zero. However, if we think for example about a single-mode squeezed vacuum, reducing the noise by a factor $e^{-2r}$ requires finely controlling $\sinh^2r$ photons~\cite{Shi:2022wpf}, so reaching zero requires  infinitely many photons. This is not necessarily the optimal way to go beyond the SQL in our context~\cite{Shi:2022wpf}, but gives an idea of the resources involved. For reference, the state of the art in fundamental physics searches corresponds to going below the SQL by a factor of $\mathcal{O}(50)$ in strain~\cite{Dixit:2020ymh} in the context of a non-demolition measurement in an empty cavity, i.e. quite different from what we would need here, where $U_{\rm in}$ is large. However we still find instructive to consider the smallest strain detectable in a quantum experiment that does not measure any noise in the quadrature of choice, since this is still not enough to probe very high frequency primordial backgrounds.
In this limit the sensitivity is purely signal-limited, i.e. we need to see at least one signal photon in the whole lifetime of the experiment
\be
P_{\rm sig}^{\rm min} = \frac{\hslash \w}{t_{\rm int}}\, .
\label{eq:Pmin}
\ee
Therefore the minimal strain in a maximally optimistic (but not necessarily feasible with current technology) quantum experiment is obtained by comparing the maximal signal power obtained integrating Eq.~\eqref{eq:Psig} with the minimal detectable power in Eq.~\eqref{eq:Pmin}. This gives
\be
h_{\rm min,\ quad}^Q &\gtrsim&\sqrt{\frac{2\pi\hslash}{U_{\rm in}t_{\rm int}}} \frac{1}{\mathcal{T}^{\rm quad}(\w_s)} \label{eq:quad_quantum}\, .
\ee
Parametrically this is the same as the classical result in Eq.~\eqref{eq:quad} for short integration times. However it is always equal or smaller than $h_{\rm min,\ quad}^C$, i.e. a quantum experiment is more sensitive than a classical one, because at long enough integration times $\Delta \w > 2\pi/t_{\rm int}$. The similarity of the two results reflects the extremely optimistic assumptions that went into them. In the classical case we are considering a single photon from quantum fluctuations as our only source of noise. In the quantum case we imagine to have prepared a quantum state that makes even this source of noise negligible in the quadrature that we are measuring, but we still need to see at least one photon from the signal during the lifetime of the experiment to claim a detection. As we discuss in Section~\ref{sec:conclusion} this is still not enough to detect primordial cosmological backgrounds at frequencies much above those probed by LVK.

Before concluding this Section we comment on a quadratic experiment based on single-photon counting~\cite{Vermeulen:2024vgl, McCuller:2022hum}. Ref.~\cite{Vermeulen:2024vgl} proposes to operate an interferometer at perfect destructive interference. If classical noise is negligible, the expected photon count at the output port in the absence of a signal is zero, while a stochastic signal predicts a non-zero expected number of photons. This gives SNR$ = t_{\rm int} \int d\w (S_h(\w)/S_n(\w))$, which is equal to that of a time-resolved signal in Eq.~\eqref{eq:meanSNR}. The result is a more favorable scaling of the minimal strain with integration time, $h_{\rm min} \sim 1/t_{\rm int}^{1/2}$ and potentially an orders-of-magnitude improvement over the traditional (homodyne) readout of interferometers that we discuss in Section~\ref{sec:transfer}.
In our Figures in Section~\ref{sec:conclusion} we do not show a ``photon counting" line because its scaling is in between the SQL result $h_{\rm min}\sim \sqrt{\hslash/U_{\rm in}}(\Delta \w/t_{\rm int})^{1/4}$ and the ultimate sensitivity at the Heisenberg limit $h_{\rm min}\sim (\hslash/U_{\rm in})(\Delta \w^3/t_{\rm int})^{1/4}$.

Eqs.~\eqref{eq:quad} and \eqref{eq:quad_quantum} show that high precision requires high energies and long integration times. Once all classical (and quantum) sources of noise have been eliminated improving the sensitivity to $h$ can only be done by increasing $U_{\rm in}$ or $t_{\rm int}$.
 
\subsection{Linear Signals}
Let us now consider a non-zero linear term, $\langle E_0(t) E_h(t) \rangle \neq 0$. This is a much larger signal, since it is linear in $h$. However, the very existence of this signal means that $\tilde E_0$ and $\tilde E_h$ have support on overlapping frequency ranges, so in the absence of a signal, the detector reads out a power that depends on $U_{\rm in} \neq 0$ in the signal region, which could be enormous compared to our $\mathcal{O}(h)$ effects. 

For this kind of signal we have the relation
\be
\frac{S_h(\w)}{S_n(\w)}=\frac{S_{\rm sig}^2(\w)}{S_{\rm noise}^2(\w)}\, ,
\ee
so that our sensitivity can be estimated from 
\be
\text{SNR}=\left(t_{\rm int} \int \frac{d\w}{2\pi} \frac{S_{\rm sig}^4(\w)}{S_{\rm noise}^4(\w)}\right)^{1/2} \simeq 1
\label{eq:SNRlin}\, .
\ee
In practice, linear detectors are often limited by shot noise, which is proportional to $\sqrt{U_{\rm in}}$. As a result, the minimal detectable power is not given by Eq.~\eqref{eq:Pnoise}, but is instead (in the limit $T\to 0$)  
\begin{equation}
S_{\rm noise}^{\rm lin}(\w)= \hslash \w \sqrt{\frac{U_{\rm in}}{\hslash \Delta \w}} = \hslash \w \sqrt{N_{\gamma}\frac{\w}{\Delta \w}}\, 
\label{eq:PminGen}
\end{equation}
where in the second equality we have related the number of photons $N_\gamma$ hitting the detector in a zero-signal experiment to the auto-correlation function of the background stored energy in the detector. In a linear experiment this quantity has support at the signal frequency and it contributes to the number of photons arriving at the detector. We can write the signal power in analogy to the quadratic case as
\be
\label{eq:Psig_lin}
S_{\rm sig}^{\rm lin}(\w)&\leq& h\, U_{\rm in}\, \mathcal{T}^{\rm lin}(\w_s)\frac{\w_s}{\Delta \w} \left[\Theta\left(\w-\w_s+\frac{\Delta \w}{2}\right)-\Theta\left(\w-\w_s-\frac{\Delta \w}{2}\right)\right]\, .
\ee
Then we use again Eq.~\eqref{eq:SNR} to write the minimal detectable strain for a linear classical detector by comparing Eq.~\eqref{eq:Psig_lin} and Eq.~\eqref{eq:PminGen} via Eq.~\eqref{eq:SNRlin} and obtain
\be
    \label{eq:hminLin}
    h_{\rm min,\; \rm lin}^C 
    &\gtrsim& \sqrt{\frac{\hslash}{U_{\rm in}}}\left(\frac{2\pi\Delta \w}{ \,t_{\rm int}}\right)^{1/4} \frac{1}{\mathcal{T}^{\rm lin}(\w_s)} \, ,
\ee
which we recognise as being the same as the quadratic detector in Eq.~\eqref{eq:quad}, up to possibly different transfer functions. This reflects the stochastic nature of the signal which averages to zero over long times $\langle h(t) \rangle =0$, so also a linear detector measures quantities proportional to $\langle h^2(t)\rangle$. If we had a signal with $\langle h(t) \rangle \neq 0$, such as a binary merger, we could use the SNR in Eq.~\eqref{eq:meanSNR} and obtain $h_{\rm min} \gtrsim \sqrt{2\pi \hslash/U_{\rm in}t_{\rm int}}$.

Note that we have defined the transfer function for a linear signal in a different way compared to the quadratic case, $h\mathcal{T}^{\rm lin}(\w)\equiv \langle \tilde U_{\rm sig}(\w)\rangle/U_{\rm in}$. The bottom line is that a much larger signal comes with a much larger, classically irreducible, source of noise and a linear detector does not outperform a quadratic detector on a primordial background of GWs.

Also in the linear case a quantum experiment need not measure the entire shot noise in Eq.~\eqref{eq:PminGen}. Given an input state with $N_\gamma$ photons, we can build a quantum experiment where the uncertainty on the strain scales as the so-called ``Heisenberg limit'' $1/N_\gamma$~\cite{Giovannetti:2011chh, RevModPhys.90.035005} rather than $1/\sqrt{N_\gamma}$ as in the classical result of Eq.~\eqref{eq:hminLin}. This requires resources well beyond current technology if applied to existing linear experiments.  It implies entangling the $N_\gamma \simeq 10^9$ photons in the laser of a typical GW interferometer, preparing a special vacuum state and keeping optical losses below $\sim 1/N_{\gamma}$~\cite{PhysRevA.88.041802}. However we find instructive to consider here also the ultimate (Heisenberg) quantum-limited detectable strain in a linear experiment
\be
h_{\rm min,\; \rm lin}^Q 
    &\gtrsim& \frac{\hslash}{U_{\rm in}} \left(\frac{2\pi\Delta \w^3}{ \,t_{\rm int}}\right)^{1/4}\frac{1}{\mathcal{T}^{\rm lin}(\w)}\, . \label{eq:lin_quantum}
\ee
This is obtained by  requiring at least one signal photon in the whole lifetime of the experiment and reproduces the famous Heisenberg limit of interferometers~\cite{Giovannetti:2011chh, RevModPhys.90.035005}. We give more details on what is required in practice to reach this limit in a companion paper~\cite{US}. Here we just use the above result to estimate the ultimate sensitivity of an interferometer and other linear detectors. Note that a quantum linear detector is exponentially more sensitive than all other detectors considered above since $U_{\rm in}$ and $t_{\rm int}$ are macroscopically large quantities prepared by the experimenters in their laboratory while $\hslash \simeq 10^{-34}$~J~s.

Reaching the Heisenberg limit requires special preparation of all aspects of the detector. As this can be impractical, if not unfeasible, with the power level employed in interferometric GW detectors, we can instead consider other quantum techniques. For example, LIGO and GEO-600, as well as a variety of other experiments targeting dark matter, inject squeezed vacuum states to improve their sensitivity beyond the SQL~\cite{Affeldt:2014rza,PhysRevX.13.041021,membersoftheLIGOScientific:2024elc}. This technique allows the uncertainty on the strain to scale, at best, as $1/N_\gamma^{3/4}$~\cite{caves1981quantum}. This corresponds to a minimum detectable strain given by
\begin{align}
    h_{\rm min, lin}^{Q,\rm sq.} \gtrsim \left(\frac{\hslash}{U_{\rm in}}\right)^{3/4} \left( \frac{2\pi\Delta \w^2}{ \,t_{\rm int}}\right)^{1/4}\frac{1}{\mathcal{T}^{\rm lin}(\w)}\, . \label{eq:lin_squeeze}
\end{align}
Current detectors do not squeeze the vacuum states sufficiently to reach this scaling with $U_{\rm in}$. Instead, their sensitivity is that of a classical linear detector (Eq.~\eqref{eq:hminLin}) multiplied by a factor $e^{-r}$ where $r$ is the ``squeeze factor''~\cite{PhysRevX.13.041021}. 

To conclude this Section, we briefly mention a special kind of linear detector. It is possible that the detector contains two (or more) sources of EM energy. A large low-frequency background $U_{\rm big}(\w_0)$ that interacts with the GW, generating the signal, and a smaller source $U_{\rm small}(\w_0\pm\w_g)$ that oscillates at higher or lower frequency and can interfere with the GW signal. An example of this kind is the haloscope in~\cite{Kontos} designed for axion detection. This detector does not have a classical parametric advantage compared to Eq.~\eqref{eq:hminLin} because its noise power scales as $\sim\sqrt{U_{\rm small}}$, but its signal power as $\sim h \sqrt{U_{\rm small} U_{\rm big}}$, giving in the end the same result for $h_{\rm min}$ that we derived in Eq.~\eqref{eq:hminLin}. Nonetheless, these setups might have a quantum advantage that we discuss in~\cite{US}.

The simple message that one can draw from our estimates is the same as in the quadratic case, once all classical (and quantum) sources of noise have been eliminated improving the sensitivity to $h$ can only be done by increasing $U_{\rm in}$ or $t_{\rm int}$. Increasing the mass of the detector can only get one to the point of saturating our estimates.

\section{Transfer Functions}\label{sec:transfer}

The last missing ingredient needed to get sensitivity estimates is a discussion of transfer functions.
Any mechanical system with resonant frequency $\w_m$ and quality factor $Q_m$, coupled to an EM readout with resonant frequency $\w_1$ and quality factor $Q$ can be described by the transfer function $\mathcal{T}_{\rm mech}$, while a direct EM conversion experiment can be described by the transfer function $\mathcal{T}_{\rm EM}$,
\be
\mathcal{T}^2_{\rm mech}(\w) &=& \frac{ \w_g^4 \w_1^4}{\left((\w_1^2-\w^2)^2+\frac{\w^2 \w_1^2}{Q^2}\right) \left((\w_m^2-\w^2_g)^2+\frac{\w^2_g \w_m^2}{Q_m^2}\right)}\, ,\nn \\
\mathcal{T}^2_{\rm EM}(\w) &=&\frac{ \w_g^2 \w^2 (\w_g L+\w_0 L+1)^2  }{\left((\w_1^2-\w^2)^2+\frac{\w^2 \w_1^2}{Q^2}\right)}\min[1, \w_g^2 L^2]\,\, ,
\label{eq:Tmaster}
\ee
where $L$ is the typical size of the detector, $\w_g$ the typical frequency of the GW that we take to have support over an interval $\Delta \w \lesssim \w_g$. $\w_0$ is the typical frequency of the EM field in the detector in absence of a GW, and we neglected $\mathcal{O}(1)$ factors encapsulating the geometry and couplings of the system.
A detailed calculation, including these factors, is presented in~\cite{US}. Here we give a heuristic derivation. Classically, one can use dimensional analysis to derive the equations for a mechanical mode $u_m$ coupled to a EM mode $e_0$, exciting a second EM mode $e_1$. We also consider a term that reproduces the direct interaction of the GW with $e_0$ and excites $e_1$. In detectors built to detect a displacement, such as interferometers or Weber bars, this second term is highly subleading. However it dominates for detectors that look for direct EM conversions of gravitons into photons. 

To set the notation more precisely, we can expand the electromagnetic field in the detector in normal modes as $\Evec(t, \vec x)=\sum_n e_n(t) \Evec_n(\vec x)$, and similarly for the magnetic field. The functions $e_n(t)$ carry the dimension of the field and encapsulate its time dependence. Similarly we expand any displacement $\U$ from equilibrium (including deformations) as $\U(t, \xv)=\sum_\alpha u_\alpha(t) \U_\alpha(\xv)$. Then a simple exercise of dimensional analysis in the proper detector frame leads to the Equations
\be
\left(\w_m^2-\w^2+i \frac{\w \w_m}{Q_m}\right) \tilde u_m(\w)&\simeq& - \frac{\w_g^2 L}{2}  \tilde h^{\rm TT}(\w)\, , \nn \\
\left(\w_1^2-\w^2+i \frac{\w \w_1}{Q}\right) \tilde e_1(\w)&\simeq& \int d\w^\prime \tilde e_0(\w-\w^\prime)\left\{ \begin{array}{l} g_m \tilde u_m(\w^\prime) \quad~~~ \text{mechanical} \\
g_e \w\, \tilde h^{\rm TT}(\w^\prime) \quad \text{electromagnetic}
\end{array}\right. \, , 
\label{eq:EOM}
\ee
where we have defined the couplings
\be
g_m &\equiv & -\frac{2\w_1^2}{L} \, , \\
\label{eq:mechCoupling}
g_e &\equiv & \w_g (1+\w_g L+\w_0 L)\min[1,\w_g L]\, ,
\label{eq:eCoupling}
\ee
and $\tilde h^{\rm TT}$ is the Fourier transform of the metric in TT gauge evaluated at the origin of our coordinate system (the center of mass of the detector). Here and in the rest of the paper we consider signals with support $\Delta \w_s \lesssim \w_g$, so that $h^{\rm TT}(t) \sim e^{i\w_g t}$. More details on the derivation of these equations are given in Appendix~\ref{app:EOM}.

From Eq.~\eqref{eq:EOM}, and the definition of the transfer functions in Eq.~\eqref{eq:Usig}, one can derive our ``master" transfer function in Eq.~\eqref{eq:Tmaster}. Note that in general what we call mechanical experiments can also have an EM conversion signal from the second term in Eq.~\eqref{eq:EOM}, but in existing detectors it is typically subleading. Therefore, we separate the cases for greater clarity. In Eq.~\eqref{eq:EOM} we took the effective mass of $u_m$ to infinity and neglected the backreaction of the EM modes on the mechanical mode, following our discussion in Section~\ref{sec:mech}, in order to get the smallest detectable strain. The role of $e_0$ is that of the background EM energy present at frequency $\w_0$ in the detector before the GW arrives, for instance LVK's laser or the energy in the LC circuit that reads out Weber bars, while $e_1$ stores the energy in the GW signal at $\w_0\pm \w_g$ and is read out by the detector. Eq.s~\eqref{eq:EOM} and~\eqref{eq:Tmaster} describe both of our detector toy models in Fig.~\ref{fig:Cartoon} and reproduce the sensitivity of the best existing GW detectors. In the following we give the most relevant examples.

Note that in most explicit transfer function examples in the literature you will not find our factor of $\min[1, \w_g^2 L^2]$ in the second term of Eq.~\eqref{eq:Tmaster}, because most detectors operate in the long wavelength limit $\w_g L \ll 1$. This factor comes from the form of the metric in the PDF~\cite{Manasse:1963zz,Misner1973,Maggiore, Berlin:2021txa, Berlin:2023grv} and we include it here to describe high-frequency detectors that operate in the short wavelength limit and obtain a parametric enhancement from the large frequency of the GW (see Section~\ref{sec:HF}).

One last aspect of our general transfer functions that is noteworthy is that the frequency $\w_s$, introduced in Section~\ref{sec:minimal}, where our sensitivity to the signal is maximal, does not necessarily coincide with the frequency of the GW $\w_g$. An example of such a setup is heterodyne detectors that look for graviton-photon conversion, whereby $\w_s = \w_0+\w_g$, as discussed in Section~\ref{sec:HF}. We are now ready to describe existing detectors and then estimate the ultimate sensitivity to primordial GWs. 

\subsection{Interferometers}
An interferometer can be described by the toy model in the left panel of Fig.~\ref{fig:Cartoon}, where the test masses in gray are its mirrors and the EM field in the readout its laser. Then the stored EM energy $U_{\rm in}$ is given by the laser at frequency $\w_L$ and one reads out the EM power approximately at the same frequency $\w_1=\w_L=\w_s-\w_g \simeq {\rm THz}$, much larger than the mechanical resonance of the mirrors $\w_m \simeq {\rm Hz}$ which is much smaller than the GW frequencies to which the detector is sensitive ($\w_g \simeq 10 \div 1000$~Hz). In this limit $\mathcal{T}_{\rm mech}$ in Eq.~\eqref{eq:Tmaster} reads
\be
\left(\mathcal{T}^{\rm LVK}\right)^2(\w_g) \simeq \frac{ \w_L^2}{\left(4\w_g^2 +\frac{\w_L^2}{Q^2}\right)} \simeq \frac{ \w_L^2 L_{\rm eff}^2}{\left(4\w_g^2 L_{\rm eff}^2 +1\right)} \, ,
\label{eq:LVK}
\ee
where we used that $Q$ is obtained from the full-width half-maximum of the Fabry-P\'erot cavities, and is given by $Q = \w_L L_{\rm eff}$. The quantity $L_{\rm eff} = L \mathcal{F}/\pi$ is the effective length of the interferometer written in terms of the true length of the arm $L$ and the finesse $\mathcal{F}$ of the Fabry-P\'erot cavities. In our limit of infinitely massive mirrors, noise is simply given by shot noise in the laser, i.e. Eq.~\eqref{eq:PminGen}. Additionally, the laser saturates the upper bound for the signal power in Eq.~\eqref{eq:Psig}. Then, by inspecting Eq.~\eqref{eq:LVK}, we find that the best sensitivity is obtained for $\w_g L_{\rm eff} \ll 1$ and we can write the smallest detectable strain from Eq.~\eqref{eq:hminLin} as
\be
h_{\rm min}^{\rm LVK} \gtrsim \sqrt{\frac{\hslash}{U_{\rm in}}} \left(\frac{2\pi\Delta \w}{ \,t_{\rm int}}\right)^{1/4} \frac{1}{\w_L L_{\rm eff}}\, ,
\label{eq:minLVK}
\ee
Numerically the transfer function gives a large enhancement of order $\w_L L_{\rm eff} \simeq 10^{11}$, but, as shown in Fig.~\ref{fig:results}, this is still not enough to reach interesting sensitivity at high frequencies. One might be tempted to increase the arm length to obtain a better sensitivity. While this is a perfectly good strategy at low frequencies that gave rise to a plethora of exciting planned and proposed detectors~\cite{Colpi:2024xhw, Sesana:2019vho,LISA:2017pwj,Ruan:2020smc,Baker:2019pnp,Kawamura:2011zz,TianQin:2020hid,Kuns:2019upi,Reitze:2019iox,Punturo:2010zz}
, it does not improve the sensitivity at arbitrarily high GW frequencies. Whenever $\w_g L_{\rm eff} > 1$ Eq.~\eqref{eq:minLVK} is no longer valid and the transfer function becomes $\mathcal{T}^{\rm LVK}\sim \w_L/\w_g \ll \w_L L_{\rm eff}$. 

As mentioned before, Eq.~\eqref{eq:minLVK} is an ultimate classical limit on the sensitivity of an interferometer. It approximately reproduces LVK's actual sensitivity at high frequency where shot noise dominates~\cite{Pitkin:2011yk}, but highly overestimates the actual sensitivity at lower frequencies where additional sources of noise must be included. In Fig.~\ref{fig:results} we plot our estimate of LIGO's sensitivity in Eq.~\eqref{eq:minLVK}, translated into an energy density using Eq.~\eqref{eq:Omin}, compared to an actual search for primordial backgrounds performed by the LIGO collaboration~\cite{Abbott_2019}. We also show constraints from the Holometer operated at Fermilab~\cite{Holometer:2016qoh} whose sensitivity can also be estiamted from Eq.~\eqref{eq:minLVK}. We find that their current sensitivity at high frequency is very close to saturating our classical limit $h_{\rm min}$.

\subsection{Resonant (Weber) Bar Detectors}
\label{sec:bars}
Resonant bars~\cite{Cerdonio_1997, PhysRevLett.18.498, PhysRevLett.20.1307,PhysRevD.68.022001, PhysRevD.76.102001, PhysRevLett.94.241101} or cavities used as resonant bars~\cite{Berlin:2023grv} target the GW excitation of a resonant mechanical mode of the detector with large $Q_m$. They can be described  by $\mathcal{T}_{\rm mech}$ in  Eq.~\eqref{eq:Tmaster} in the limit $\w_m\simeq \w_1 \simeq \w_g = \w_s$, where
\be
(\mathcal{T}^{\rm Weber})^2 \lesssim Q^2 Q_m^2\, . \label{eq:Weber}
\ee
To get the best sensitivity of a Weber bar we cannot simply substitute the upper bound in Eq.~\eqref{eq:Weber} into our minimal detectable strains because our minimal strains were derived using the largest possible signal power at a given frequency $P_{\rm sig} \simeq \w_s U_{\rm sig}$. The upper bound on $\mathcal{T}^{\rm Weber}$ is only saturated in a small window around the resonant frequency, so the signal power is given by $P_{\rm sig} \sim \Delta \w\, U_{\rm sig} \sim (\w_s/Q) U_{\rm sig} \ll P_{\rm max} \sim \w_s U_{\rm sig}$. 
Therefore we use the following PSD to obtain the sensitivity of a quadratic-in-strain resonant experiment 
\be
S_{\rm sig}^{\rm res}(\w)&=&h^2 U_{\rm in}\left(\mathcal{T}^{\rm quad}(\w_s)\right)^2\left[\Theta\left(\w-\w_s+\frac{\Delta \w}{2}\right)-\Theta\left(\w-\w_s-\frac{\Delta \w}{2}\right)\right]\, , \label{eq:Pres}
\ee
with $\Delta \w$ that depends on the details of the experiment. If the experiment operates on both an EM and a mechanical resonance, $\w_s=\w_g \simeq \w_m \simeq \w_1$, we assume a signal broader than the detector bandwidth $\Delta \w_s > \Delta \w_d$ and a long enough integration time $\Delta \w > 2\pi t_{\rm int}^{-1}$, then $\Delta \w_{_{\rm Weber}}= \w_g/\max[Q,Q_m]$.
Therefore, for a Weber bar on resonance, we obtain
\be
h_{\rm min}^{\rm Weber} \gtrsim \sqrt{\frac{\hslash}{U_{\rm in}}}\left( \frac{2\pi\,\w_g \text{max}[Q,Q_m]}{t_{\rm int}}\right)^{1/4} \frac{1}{Q_m Q}\, .
\label{eq:hminWeber}
\ee
While in principle the signal transfer function could be larger than LVK, $Q_m Q \lesssim 10^{12}$, this optimal Weber bar would be narrowband. Furthermore, it is almost impossible to realise this optimum in practice.

One should recall that when searching for GW signals on the mechanical resonance, the limiting noise is typically due to displacement noise in the mechanical resonator. As such, our estimate above is far too optimistic, as it assumes EM noise dominates over mechanical noise. As a case study of when EM noise dominates over mechanical noise, we can consider the recently proposed Magnetic Weber Bar (MWB)~\cite{Domcke:2024mfu}. In that proposal, on the mechanical resonance, thermal mechanical noise is over $10^6$ times larger than EM readout noise. In order to reduce the mechanical thermal noise to the level of SQUID noise, one would have to increase the magnet mass to over $M \gtrsim 10^{12}\,\text{kg}$, i.e., the mass of a small asteroid. An object of density $\rho \sim 6\,\text{g/cm}^3$ would occupy a volume of $V\sim (10^3\,\text{m})^3$ to have this mass, demonstrating the difficulty of achieving the optimistic sensitivity scaling given in Eq.~\eqref{eq:hminWeber}. 

This is clearly an extreme requirement on the mass of a resonator for EM noise to be dominant, so for completeness we also compute the scaling of the minimum detectable strain on the mechanical resonance, including mechanical noise. This is given by identifying $U_{\rm in}$ with the mechanical energy $U_{\rm mech} = \frac{1}{2}M \w_{m}^2 L^2$, and replacing $Q\to 1$, since the EM resonance filters both signal and noise identically, yielding
\begin{align}
    \left(h_{\rm min}^{\rm Weber}\right)_{\rm mech} \gtrsim \sqrt{\frac{\hslash}{U_{\rm mech}}}\left( \frac{2\pi\,\w_g\, Q_m}{t_{\rm int}}\right)^{1/4}\frac{1}{Q_m} \ .
    \label{eq:hminWeberMech}
\end{align}
In practice, the minimum sensitivity will therefore be given by the greater of Eqs.~\eqref{eq:hminWeber} and~\eqref{eq:hminWeberMech}. If we instead consider the sensitivity away from a mechanical resonance, but still on an EM resonance, then the part of Eq.~\eqref{eq:Tmaster} due to mechanical displacements is given by $\text{min}[1,(\wg/\w_m)^4]$ instead of $Q_m^2$. Therefore the minimum detectable strain in the infinite mass limit is given by
\begin{align}
\label{eq:WeberBroad}
    \left(h_{\rm min}^{\rm Weber}\right)_{\rm non-res} \gtrsim h_{\rm min}^{\rm Weber} \times \frac{Q_m}{ \text{min}[1,(\wg/\w_m)^2]} \ ,
\end{align}
with the bandwidth factor in Eq.~\eqref{eq:hminWeber} set purely by $Q$ and not $Q_m$.
This scaling demonstrates the large penalty in sensitivity, $(\w_m/\wg)^2$, incurred when searching for signals below a mechanical resonance, i.e., in the \emph{rigid detector} regime. 

Finally, if we consider a non-resonant EM readout, the minimum detectable strain sensitivity is further degraded by a factor of $Q^{3/4}/\text{min}[1,(\w_1/\wg)^2]$. This is the regime in which the MWB proposal operates~\cite{Domcke:2024mfu}. Although naively this would seem to be an unwanted penalty, it has the benefit of allowing for a potentially much larger bandwidth of $\Delta \w \sim \w_s = \wg$. The figure of merit for cosmological GW detection is the combination $h_{\rm min}^2/\Delta \w$. We show this in Eq.~\eqref{eq:Omin} and Appendix~\ref{app:energy} when discussing stochastic backgrounds of GWs characterized by their energy density. Comparing the different scenarios, we find that the figure of merit is given by 
\begin{align}
\label{eq:mech}
\frac{h_{\rm min}^2}{\Delta \w} = \frac{\hslash}{U_{\rm in}}\sqrt{\frac{2\pi}{t_{\rm int}\, \wg}} \times
    \begin{cases}
        \frac{\text{max}[Q_m,Q]^{3/2}}{(Q_m Q)^2} \ , ~~~ &\text{Optimal resonant,} \\
        \frac{U_{\rm in}}{U_{\rm mech}}\frac{1}{\sqrt{Q_m}} \ ,~~~&\text{Mech noise dominated,} \\
        \frac{1}{\text{min}[1,(\wg/\w_m)^4]} \frac{1}{\text{min}[1,(\w_1/\wg)^4]} \ ,~~~&\text{Broadband.}
    \end{cases}
\end{align}
The optimal resonant experiment in Eq.~\eqref{eq:mech} is the one that saturates the upper bound in Eq.~\eqref{eq:Weber} sitting on top of a mechanical and a EM resonance and does not have mechanical noise. The mechanical noise dominated expression corresponds to the case that is more likely to occur on a mechanical resonance: mechanical noise exceeds EM noise, so the EM resonance no longer improves the sensitivity, which is in turn dictated by the mechanical energy instead of the EM energy.
The broadband search is performed away from any resonance, but still in absence of mechanical noise.
The result is that at very high frequencies, the optimal resonant sensitivity to cosmogenic GWs can be inferior to that of a broadband experiment.

To make this discussion more concrete we consider the existing Weber bar detector AURIGA~\cite{Cerdonio_1997}. Its sensitivity is limited by mechanical noise as in Eq.~\eqref{eq:hminWeberMech}. The mechanical mode used in~\cite{Branca:2016rez} has $\w_m \simeq 2\pi \times 900$~Hz, $M\simeq 1.1\times 10^3$~kg, $L\simeq 3$~m and $Q_m\simeq 10^6$ and gives $h_{\rm AURIGA}^{\rm mech} \simeq 10^{-26} (s/t_{\rm int})^{1/4}$. A setup purely limited by its LC circuit in the readout ($Q\simeq 10^6$, $U_{\rm in}\simeq 2$~kJ~\cite{Bonaldi:1998gcg}) would give $h_{\rm AURIGA}^{\rm EM} \simeq 10^{-2} h_{\rm AURIGA}^{\rm mech}$. As expected, our optimistic assumptions largely overestimate the actual sensitivity $\sqrt{S_h} \simeq 10^{-21}$~Hz$^{-1/2}$~\cite{Branca:2016rez}. When comparing our $h_{\rm min}$ with $\sqrt{S_h}$ quoted by experiments it is useful to keep in mind that $\sqrt{S_h}$ is typically their noise budget, while our $h_{\rm min}$ refers to a hypothetical signal which is constant over $\Delta \w$. So $h_{\rm min} \simeq \sqrt{2\pi S_h \Delta \w}/\sqrt{t_{\rm int}\Delta \w}=\sqrt{2\pi S_h/t_{\rm int}}$.

It is especially instructive to compare AURIGA's best possible sensitivity to LVK's minimal estimated strain in the previous Section, which is roughly $h_{\rm min}^{\rm LVK} \simeq 10^{-23} (s/t_{\rm int})^{1/2}$. This does not mean that Weber bars are better GW detectors than interferometers, but rather that they would become better if one could realize in practice our extremely optimistic assumptions on noise. To get $h_{\rm AURIGA}^{\rm mech} \simeq 10^{-26} (s/t_{\rm int})^{1/4}$ we assumed a single quantum of mechanical noise at the signal frequency, i.e. $U_{\rm mech}^{\rm noise} = \hslash \w_g$, which is far from realized in practice. We can recover AURIGA's sensitivity by noticing that the apparatus is cooled to liquid Helium temperatures. This implies from Eq.~\eqref{eq:Pnoise} about $10^6$ phonons of noise. On the other hand existing interferometers almost saturate our optimistic estimates. One should keep in mind these assumptions when reading the plots in Section~\ref{sec:conclusion}, where our ``crazy" resonator looks competitive with our ``crazy" interferometer.

\subsection{Optomechanical Sensors}
\label{sec:optomech}
Optically levitated sensors~\cite{Aggarwal:2020umq, Arvanitaki:2012cn} are small dielectric objects of typical size $\sim \mu $m that are kept in equilibrium by a laser. Another laser cools down the mechanical mode corresponding to their center of mass displacement and reads out its position. Their small size is compensated by very high mechanical quality factors on resonance. The parametric scaling of their sensitivity at high frequency can look promising compared to interferometers~\cite{Aggarwal:2020umq}.

It is easy to express their sensitivity in our formalism. The readout laser has the largest frequency in the problem $\w_1 \simeq \w_L \gg \w_g$, as in the case of interferometers. However, a mechanical mode of the sensor can be resonantly excited by GWs $\w_g\simeq \w_m$, as we discussed for Weber bars. Therefore their signal transfer function $\mathcal{T}_{\rm mech}$ on a mechanical resonance reads
\be
\left(\mathcal{T}^{\rm opto}\right)^2(\w_g)= \frac{ Q_m^2 \w_L^2}{\left(4\w_g^2 +\frac{\w_L^2}{Q^2}\right)} \ .
\ee
As in the case of LVK, the quantity $Q$ is related to the finesse of the cavity and the length by $Q = \w_L L \mathcal{F}/\pi = \w_L L_{\rm eff}$, with typical lengths $L_{\rm eff} \simeq 10^{-3} L_{\rm eff}^{\rm LVK}$.
 Using Eq.~\eqref{eq:Pnoise} for a detector sensitive to an observable that is linear in strain, the result is that the optimal sensitivity of optomechanical sensors is approximately
\begin{align}
    h_{\rm min}^{\rm opto} \gtrsim \sqrt{\frac{ \hslash}{U_{\rm in} }}\left(\frac{2\pi\Delta \w}{ \,t_{\rm int}}\right)^{1/4} \frac{1}{Q_m \w_L L_{\rm eff}} \ ,
    \label{eq:minOpto}
\end{align}
in the $\wg L_{\rm eff} \ll 1$ limit. 
If we consider the experimental setup proposed in~\cite{Arvanitaki:2012cn} for a levitated sphere, we have a readout laser with $P_{\rm in}=1.1$~W and $\lambda_L=1.55\,\mu$m, a mechanical quality factor $Q_m\simeq 10^{6}$ and a cavity with $\mathcal{F}=10$ and $L=10$~m. This gives a rather spectacular $h_{\rm min}^{\rm opto} \simeq 10^{-22} (s/t_{\rm int})^{1/4}$ best possible sensitivity on resonance. However, as was the case for Weber bars, the actual sensitivity is degraded, $h_{\rm min} \simeq 7\times 10^{-17}/\sqrt{\rm Hz}$ due to large sources of noise (mainly the thermal motion of the sphere center of mass) that we did not include in Eq.~\eqref{eq:minOpto}. Our estimate corresponds to having a photon shot-noise limited displacement sensitivity $S_{\delta x}(\w)\simeq (\lambda_L/\mathcal{F})\sqrt{\hslash \w_L/P_{\rm in}}$ as our only source of noise~\cite{Anetsberger_2009}.

\subsection{Electromagnetic Conversion}\label{sec:HF}

In the case of electromagnetic conversion experiments, we can use the second term $\mathcal{T}_{\rm EM}$ in the signal transfer function of Eq.~\eqref{eq:Tmaster} to estimate the minimum detectable strain of different types of detectors. Broadly speaking, these experiments fall in one of four categories: conversion in a DC background EM field ($\w_0=0$) either off or on an EM resonance, and conversion in an AC background EM field off or on an EM resonance. At the end of this Section we compare these different scenarios and identify the most sensitive setups.

\subsubsection*{DC background field, off EM resonance}
In this case we have different parametric scalings depending on the hierarchy between $\w_g, \w_1$ and $L$. Not surprisingly, the worst case is when $\w_g$ is the smallest scale in the problem. If $\wg \ll \w_{1} \simeq 1/L$ the signal transfer function-squared is given by $(\mathcal{T}^{\rm DC})^2 \simeq (\wg L)^6$. 

An off-resonant DC conversion experiment where $\w_g$ is the largest scale in the problem, i.e. $\wg \gg \w_1 \simeq 1/L$ has much more favourable scaling. In this case the transfer function-squared is $(\mathcal{T}_{\rm sig})^2 \simeq (\wg L)^2 \gg 1$. If we consider a broad signal, $\Delta \w \simeq \w_g$, and a sufficiently long integration time, the minimum detectable strain is then given by 
\begin{align}
    h_{\rm min}^{\rm DC,~non-res.} \gtrsim \sqrt\frac{\hslash}{{U_{\rm in}}}
    \left(\frac{2\pi\, \wg}{t_{\rm int}}\right)^{1/4}
    \frac{1}{(\wg L)} \ , \quad \wg L \gg 1\ .
    \label{eq:DCnonResAbove}
\end{align}
Examples of such an experiment include MADMAX~\cite{Domcke:2024eti}, CAST or IAXO~\cite{Ejlli:2019bqj}. These all have similar geometries with one long and two short axes, with the photodetector placed at the end of the long axis. Their readout is not resonant (i.e. $Q\simeq 1$ in our language).

\subsubsection*{DC background field, on EM resonance} 

Typically, a high-$Q$ resonance also implies that $\wg L \simeq 1$, as higher-order modes for which $\wg L >1$ will have lower $Q$. As such, we find that the signal transfer function is given by $(\mathcal{T}^{\rm res})^2 \simeq Q^2$. The result is the same as for a Weber bar with $Q_m=1$, as it is obtained from the same form of the signal PSD (Eq.~\eqref{eq:Pres}),
\be
h_{\rm min}^{\rm DC,~res} \gtrsim \sqrt\frac{\hslash}{{U_{\rm in}}}
\left(\frac{2\pi\,\w_g}{t_{\rm int}}\right)^{1/4}
\frac{1}{Q^{3/4}}\, .
\label{eq:DCRes}
\ee
To obtain this result we assumed a long integration time compared to the inverse resonator bandwith, i.e. $\Delta \w = \wg/Q \ll 2\pi/t_{\rm int}$.

Examples of such detectors include resonant EM cavities as considered in Ref.~\cite{Berlin:2021txa}, or lumped LC resonators as considered in Ref.~\cite{Domcke:2022rgu}. In the latter case, $\wg L$ need not be order unity, as the resonance condition depends on the LC circuit parameters and not necessarily on the physical dimensions of the detector. If $\w_g L \gg 1$ the sensitivity is further enhanced by a factor of $\w_g L$ in the transfer function.  However in practice it is non-trivial to build large LC resonators with high quality factors. 

\subsubsection*{AC background field, below EM resonance}
When considering AC background fields, we will assume a quadratic-in-strain measurement. Note that in certain circumstances, the linear signal will also exist, in which case our sensitivity estimates are not quite the optimal ones. However, as discussed in Section~\ref{sec:minimal}, the main difference between linear and quadratic signals appears only in the regime where quantum techniques are applied. 
It should also be recalled that DC fields in the laboratory are typically much larger than their AC counterparts, so the stored EM energy is much larger for a DC experiment. 

If we take $\w_0\simeq \w_1$, and $\wg \ll \w_{0,1}$, the transfer function is approximately 
\begin{align}
    \left(\mathcal{T}^{\rm AC,~non-res.}\right)^2 \simeq \frac{Q^2 \w_g^2(1+ \w_0 L)^2\text{min}[1,(\wg L)^2]}{\w_0^2 + 4 Q^2 \wg^2} \ .
\end{align}
The signal frequency is $\w_s = \w_0 + \wg \sim \w_0$.
We see that in the PDF, if $\w_0 \simeq 1/L$, the dominant contribution to the signal is simply $\left(\mathcal{T}^{\rm AC,~non-res.}\right)^2 \sim (\w_g/\w_0)^2\text{\rm min}[Q^2 (\w_g/\w_0)^2,1]$. The resulting optimal sensitivity is given by
\begin{align}
    h_{\rm min}^{\rm AC,~non-res.} \gtrsim \sqrt\frac{\hslash}{{U_{\rm in}}}\left(\frac{2\pi\w_g}{t_{\rm int}}\right)^{1/4} \frac{1}{(\w_g/\w_0)\text{min}[Q (\w_g/\w_0), 1]} \ .
\end{align}
This case corresponds to the broadband operation mode of MAGO 2.0 reading out only its EM signal~\cite{Berlin:2023grv}. For concreteness, in that proposal a cavity of mass $M=10\,\text{kg}$ is loaded with an AC EM field oscillating at $\w_0 \sim 2\pi \cdot\text{GHz}$, corresponding to a stored EM energy of $\mathcal{O}(100)\,\text{J}$. The intrinsic EM quality factor of the cavity is $Q = 10^{10}$, although optimal operation requires overcoupling so that the effective quality factor is $Q \sim 10^5$. We do not show the case where $\w_g$ is the largest scale in the problem because the transfer function has the same scaling as the DC case, but the stored energy is smaller.

\subsubsection*{AC background field, on EM resonance}

When the resonance condition is reached, i.e. $\w_1 = \w_0+\wg$, we find that the transfer function is approximately 
\begin{align}
(\mathcal{T}^{\rm AC,~res})^2 \simeq \frac{Q^2 \w_g^2}{\w_1^2}(\wg L+\w_0 L +1)^2\text{min}[1,(\wg L)^2] \ .
\end{align}
Note that the factor $\wg L$ need not be order unity when converting in an AC background field, as the resonance is associated to $\w_s = \w_1= \w_0 + \wg$. The result is that $\w_s \sim a/L$, where $a$ is a numerical constant. In an RF cavity it is likely to be $a = \mathcal{O}(1)$, while in an optical cavity it can be much larger. As a result, we find that the optimal sensitivity is
\begin{align}
    h_{\rm min}^{\rm AC,~res} \gtrsim \sqrt\frac{\hslash}{{U_{\rm in}}}\left(\frac{2\pi\w_1}{Qt_{\rm int}}\right)^{1/4} \frac{1}{Q(\wg/\w_1)(\w_1 L + 1)\text{min}[1, \wg L]} \ ,
\end{align}
where we recall that $\w_1 = \w_0 + \wg$. This corresponds, for example, to resonant operation of MAGO 2.0 for an electromagnetic signal~\cite{Berlin:2023grv}. In that analysis, only the regime where $\wg L \ll 1$ was considered. However, on the basis of the scaling above, it would suggest that resonant operation in the regime $\wg L \gg  1$ could lead to an improved sensitivity by a factor of $\wg L$, but we leave to future work the description of a detector geometry that can actually take advantage of this factor.

\subsubsection*{Cross-comparison of EM conversion}

To summarise our findings for the various regimes in which an EM conversion experiment can operate when searching for GWs, we examine the figure of merit for cosmological sources. As discussed in the context of Weber bars, and will be made more apparent in the next section, the figure of merit is $h_{\rm min}^2/\Delta \w$. For the five scenarios considered above, we obtain
\begin{align}
    \frac{h_{\rm min}^2}{\Delta \w} \gtrsim
    \sqrt{\frac{2\pi}{t_{\rm int}}}\frac{\hslash}{U_{\rm in}} \times 
    \begin{cases}
        \frac{1}{\sqrt{\wg}(\wg L)^2} \ , \quad \text{DC, non-res.,~}\\
        \frac{1}{\sqrt{\wg}}\frac{1}{\sqrt{Q}}\ , \quad \text{DC, res.} \\
        \frac{1}{\sqrt{\w_g}}\frac{1}{(\w_g/\w_0)^2\text{min}[Q^2 (\w_g/\w_0)^2, 1]} \ , \quad \text{AC, non-res.,}\, \wg \ll \w_0 \\ 
        \sqrt{\frac{Q}{\w_g}}\frac{1}{(Q(\wg/\w_1)(\w_1 L + 1)\text{min}[1, \wg L])^2}\ , \quad \text{AC, res.}
    \end{cases}
\end{align}
In the above, we have assumed that the bandwidth is given by $\Delta\w = \wg$ in the case of DC non-resonant experiments, by $\Delta \w = \wg/Q$ for DC resonant experiments, $\Delta \w = \w_g$ for AC non-resonant and finally $\Delta \w = (\w_0 + \wg)/Q = \w_1/Q$ for AC resonant searches. We see the intrinsic advantage of AC searches lies in the potentially large hierarchy between $\w_0$ and $\wg$. However, this is negated if $\w_0/\wg > Q$. Furthermore, it should be recalled that AC experiments can often have noise sources beyond the minimum we have considered here. 

In summary, the optimal sensitivity for a DC EM field experiment is obtained on an EM resonance if $\wg L \ll 1$, and potentially off-resonance if $\wg L \gg Q^{1/4}$. For AC field experiments, we see that the optimal sensitivity is obtained on the EM resonance. However, this comes at the cost of $U_{\rm in}^{\rm AC} \ll U_{\rm in}^{\rm DC}$. 
Finally, we recall that above we have considered AC field experiments operated in the quadratic-in-strain regime. If the linear signal is also present, it could present an advantage for an AC field experiment when utilising quantum resources.

\section{Cosmological Backgrounds}\label{sec:signals}
We would like to convert the minimal detectable strains in the previous Sections into a sensitivity to cosmological gravitational radiation. A primordial stochastic background of GWs can be modeled as a gaussian, stationary, isotropic and unpolarized random process with $\langle h(t) \rangle =0$~\cite{Maggiore_2007}. Therefore the signal that we are looking for starts at order $h^2$ and we can characterize it in terms of its energy density $\rho_g(\w) \propto \langle \dot h^2(t)\rangle$. It is common to define the relative energy density in GWs as
\be
\Omega_g(\w) \equiv \frac{8\pi G_N}{3 H_0^2}\frac{d \rho_g(\w)}{d\log \w}\, ,
\ee
where $H_0$ is the value of the Hubble constant today. Following the derivation in Appendix~\ref{app:energy}, we can write the minimal detectable energy density for a classical detector (i.e. a detector operating at its SQL, where all classical sources of noise have been made negligible)
\be
   \label{eq:OmegaGgen}
   \big( \Omega_g(\w) \big)_{\rm min}^C 
   &\simeq & \mathcal{N}\frac{\w_s^3}{3 H_0^2} \frac{h_{\rm min}^2}{ \Delta \w} \gtrsim  \mathcal{N}\frac{\w_s^2}{3 H_0^2}\frac{\hslash\w_s}{U_{\rm in}} \sqrt{\frac{2\pi}{t_{\rm int} \Delta \w}}\frac{1}{\mathcal{T}^2(\w)} \ . 
   \label{eq:Omin}
\ee 
Here $\mathcal{N}$ is a $\mathcal{O}(1)$ number that depends on the detection scheme and we neglect in the following. To derive Eq.~\eqref{eq:Omin} we assumed a constant energy spectrum in the interval $\Delta \w$. We see that the sensitivity to stochastic GW backgrounds scales as the inverse of the supplied EM energy, and will vary from detector to detector primarily through the transfer function differences. We can similarly derive a lower bound for a quantum detector, operating beyond its SQL,
\be
\big( \Omega_g(\w) \big)_{\rm min}^Q 
\gtrsim  \mathcal{N}\frac{\w_s^2}{3 H_0^2} \frac{\hslash \w_s}{U_{\rm in}}\left\{\begin{array}{c}\frac{2\pi}{\Delta \w t_{\rm int}}\left(\frac{1}{\mathcal{T}^{\rm quad}(\w)}\right)^2\quad \text{quadratic} \\ \frac{\hslash}{U_{\rm in}}\sqrt{\frac{2\pi\Delta \w}{t_{\rm int}}}\left(\frac{1}{\mathcal{T}^{\rm lin}(\w)}\right)^2\quad\text{linear}\end{array}\right.
. \label{eq:OminQ}
\ee
The quantum linear result is orders of magnitude better  than its classical counterpart, as it is enhanced by the huge factor $U_{\rm in} t_{\rm int}/\hslash$. However, one should recall the extremely optimistic that went into deriving Eq.~\eqref{eq:OminQ}, which minimally requires manipulating all $U_{\rm in}/\hslash \w_L$ photons into a specific quantum state. Nonetheless we will show that even current interferometers operated at the Heisenberg limit, i.e. Eq.~\eqref{eq:OminQ}, cannot detect very high frequency primordial backgrounds.

Our primary physics target will be the BBN/CMB bound on primordial sources of energy density. A primordial background of GWs detectable in the laboratory today must have an energy density that respects the constraint~\cite{Kawasaki:1999na,Kawasaki:2000en,Hannestad:2004px, Planck:2018vyg}
\be
\int d\log \w \; h^2_{\rm eff}\Omega_g(\w) \lesssim 5\times 10^{-6} \Delta N_{\rm eff}\, , \label{eq:BBN}
\ee
where $h^2_{\rm eff}\simeq 0.68$ is the reduced Hubble parameter and $\Delta N_{\rm eff} \lesssim 0.2$~\cite{Mangano:2011ar, Cyburt:2015mya, Peimbert:2016bdg, Planck:2018jri} is the uncertainty on the effective number of neutrino species. We use it as a (generous) benchmark for primordial signals, that in practice are often much smaller. However, examples of plausible high-frequency signals saturating the bound exist (see for example~\cite{Servant:2023tua}) and we display two of them in the next Section.

\begin{figure}[t]
\centering
\includegraphics[scale=0.9]{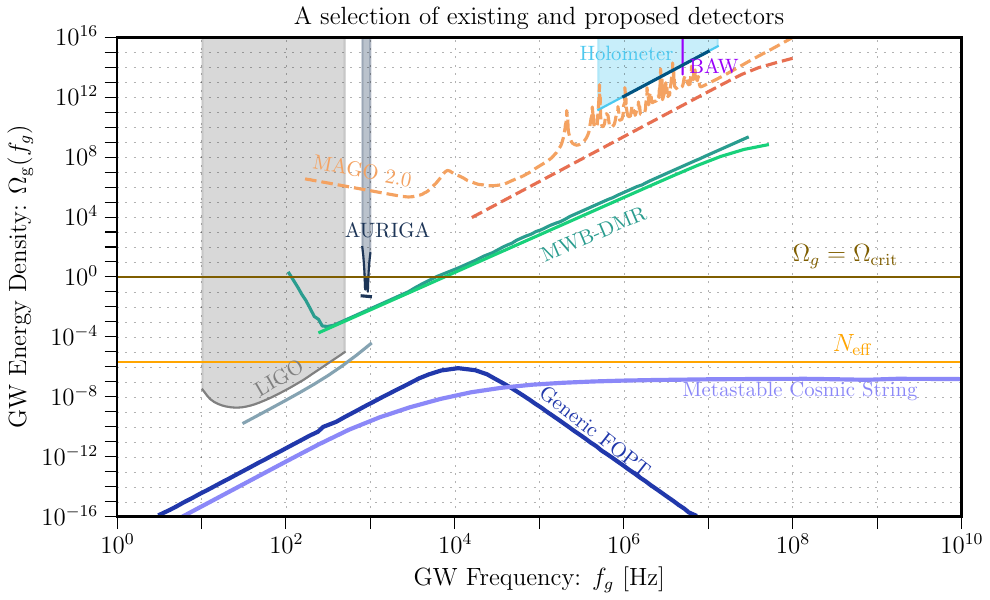}
\caption{Sensitivity to a stochastic background of GWs. The existing limit from LIGO is shown as a solid gray line, with our heuristic estimate shown as a thicker gray line beneath it. The expected design sensitivity of MAGO is shown as a dashed orange line, with our heuristic estimate shown as a dashed red line. The expected sensitivity of a Magnetic Weber Bar based on the DMRadio-GUT magnet is shown as a dark green line, with our heuristic estimate shown as a lighter green line. The Holometer sensitivity in light blue is compared to our estimate in darker blue. For LIGO and the Holometer we show their searches for a specific stochastic primordial background that are detailed in~\cite{Abbott_2019,Holometer:2016qoh}. For comparison, we show two hypothetical BSM signals that approximately saturate the BBN bound (gold), namely a generic first-order phase transition (FOPT) in blue and metastable cosmic strings in purple. More details on the signals can be found in~\cite{Servant:2023tua}.}
\label{fig:results}
\end{figure}

\begin{figure}[t]
\centering
\includegraphics[scale=0.9]{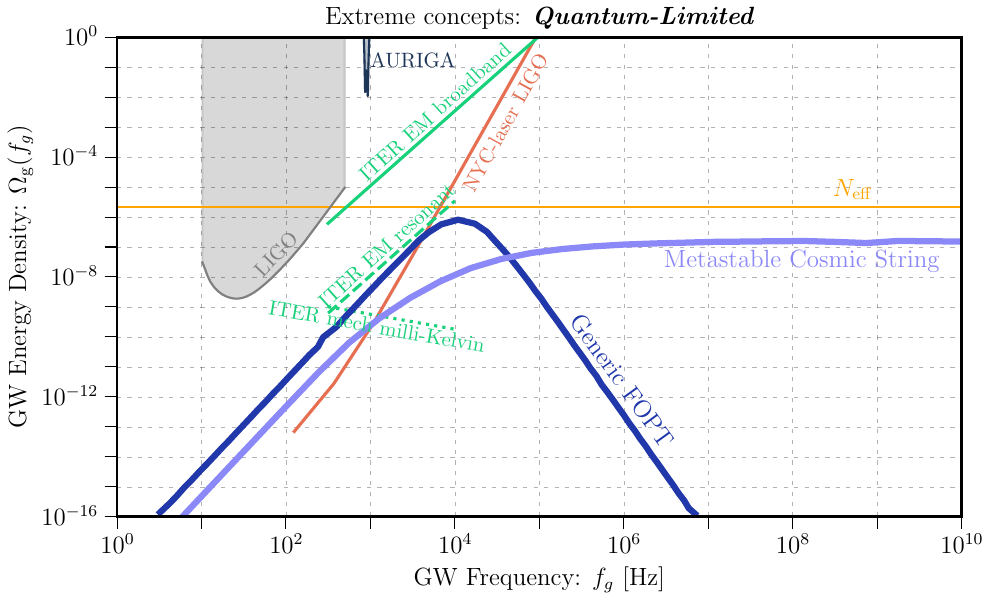}
\caption{Projected sensitivity of utterly implausible detectors operated at the classical limits of quantum noise. 
We show in orange a LIGO-sized interferometer operated with a laser power of $10^{10}\,\text{W}$ (comparable to the maximum power usage of NYC), limited only by shot-noise detecting a signal with $\Delta \w=2\pi\times1$~kHz over a data taking period of one year. In principle, backaction noise limits the sensitivity severely for such high power. In order for the NYC-laser LIGO to probe beyond the $N_{\rm eff}$ bound (yellow line), a mirror mass in the range $10^{8}\gtrsim M/\text{kg}\gtrsim 10^2$ is required, where the range applies to the frequency range $10^2\lesssim f_g/\text{Hz} \lesssim 10^4$.
In green we show three different EM/mechanical detectors, taking the approximate parameters of the ITER fusion reactor: $U_{\rm mech} \sim 10^{16}\,\text{J}\times (f_g/\text{kHz})^2$, $U_{\rm in} = 10^{12}\,\text{J}$. The solid green line assumes the ITER magnetised volume ($11.8\,\text{T},~8500 \text{m}^3)$ operated as a Magnetic Weber Bar with a broadband detection scheme. The dashed green line assumes that at every frequency there is an EM resonator with a quality factor $Q=10^6$ to enhance the sensitivity. The dotted green line assumes mechanical energy dominates on a mechanical resonance, and noise is limited by a bar temperature of $T = 1\,\text{mK}$ with $Q_{m} = 10^6$. While this is not technically quantum-limited, as $T/\w \gtrsim 10^3$ across the frequency range, so to reach the single phonon limit would require cooling to $\mu\text{K}$ or lower temperatures. All three lines assume $t_{\rm int} = \text{yr}$. In the case of the resonant EM and mechanical lines, this integration time is taken \emph{at each frequency}, so that scanning the entire range shown would take about $t_{\rm total} \sim 4\times 10^6\,\text{yr}$.
}
\label{fig:crazy_classical}
\end{figure}

\begin{figure}[t]
\centering
\includegraphics[scale=0.9]{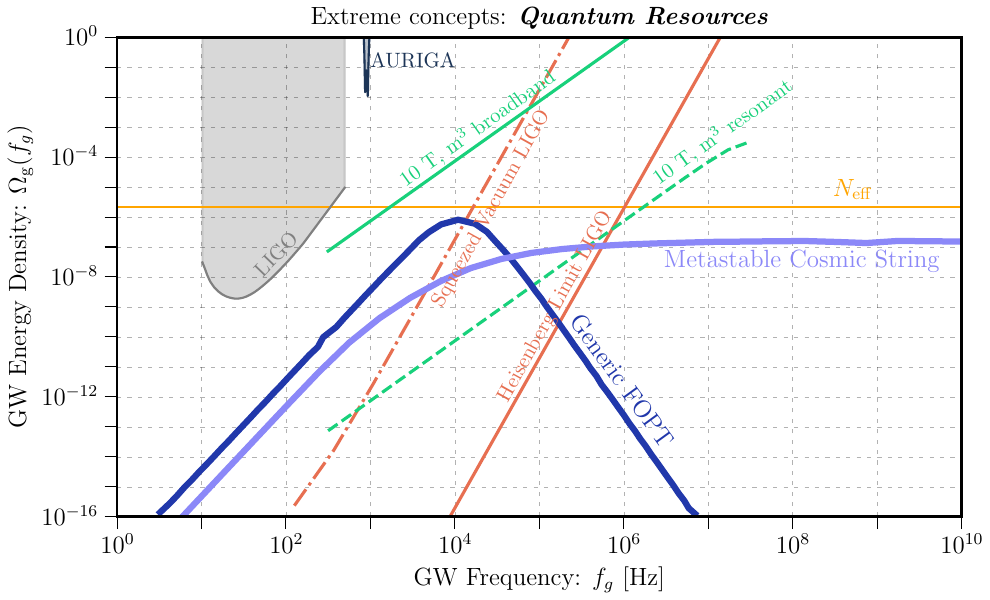}
\caption{
We show implausible detector concepts that go beyond classical detection schemes, and employ (significant) quantum resources to achieve. In orange, we show $P_{\rm in} = \text{kW}$ lasers in an interferometer experiment that is otherwise LIGO-like. The solid line shows the sensitivity of such a device if operated at the Heisenberg limit. Achieving this requires manipulation and control of all $N_\gamma$ quanta in the laser, and reduction of losses to much less than $1/N_\gamma$. The dot-dashed orange line shows the sensitivity of a device where the laser is not directly manipulated, and instead the vacuum is squeezed to the maximum theoretical limit~\cite{caves1981quantum}. In solid and dashed green we show the sensitivities that could be achieved with Magnetic Weber Bars with a magnetised volume of a cubic metre and a magnetic field of $10\,\text{T}$. For the resonant version, we assume $Q=10^6$. 
These green curves assume that the noise is reduced to the point where a single photon of signal is detectable in the entire duration of the experiment. All curves assume an integration time of a year.
}
\label{fig:crazy_quantum}
\end{figure}

\section{Results and Concluding Remarks}\label{sec:conclusion}
We show our main results in Figures~\ref{fig:results}, \ref{fig:crazy_classical} and~\ref{fig:crazy_quantum}. In Fig.~\ref{fig:results} we display a selection of existing and proposed detectors. We plot both the actual sensitivity as provided in the literature, and our estimates using Eq.~\eqref{eq:Omin} with our Eqs.~\eqref{eq:quad},~\eqref{eq:hminLin} and the transfer functions given in Section~\ref{sec:transfer}. Our estimates match or exceed the actual sensitivity of a given detector, as expected given the optimistic assumptions we made in their derivation. In addition to the existing sensitivities of LIGO~\cite{Abbott_2019} (grey), AURIGA~\cite{Vinante:2006uk} (dark gray), the Holometer~\cite{Holometer:2016qoh} (light blue) and Bulk Acoustic Wave resonators~\cite{Goryachev:2021zzn} (purple), we show the expected sensitivities of proposed experiments such as Magnetic Weber Bar-DM Radio (MWB-DMR)~\cite{Domcke:2024mfu} (dark green) and MAGO 2.0~\cite{Berlin:2023grv} (orange).

To demonstrate that our calculations of sensitivities and transfer functions give a reasonable result when compared with a more careful analysis, we also show the sensitivity curves that one would obtain using our Eqs.~\eqref{eq:quad},~\eqref{eq:hminLin} and~\eqref{eq:Omin} with the appropriate transfer functions. In the case of LIGO, we use the laser parameters described in Ref.~\cite{PhysRevX.13.041021} in our Eq.~\eqref{eq:minLVK} to estimate the sensitivity after one year of integration with a bandwidth of $\Delta \w = 2\pi\times 25\,\text{Hz}$~\cite{Abbott_2019}. This is shown as a light grey solid line that sits below the true LIGO bound of Ref.~\cite{Abbott_2019}. Our estimate is a bit optimistic by design, reflecting the major assumptions we made about signal and noise in deriving Eq.~\eqref{eq:minLVK}. In particular we took the signal to be maximal and constant over the whole bandwidth $\Delta \w$, while~\cite{Abbott_2019} is looking for a energy density scaling as a power-law.

We also show our estimate of the Holometer (another laser interferometer) sensitivity as a blue line, which precisely matches the true Holometer bound of Ref.~\cite{Holometer:2016qoh} (light blue shaded region). We use the parameters described in the same reference for our estimate, i.e. an input laser power of $P_{\rm in} = \w_L U_{\rm in} = 1\,\text{W}$ in an interferometer of length $L = 39\,\text{m}$ and finesse $\mathcal{F} = 2000$. The laser wavelength is $\lambda_L = 1064\,\text{nm}$.

We show the sensitivity of AURIGA (dark grey shaded region), which we computed using the noise-equivalent strain PSD given in Ref.~\cite{Branca:2016rez} and assumed their 10-yr dataset and a bin size of $\Delta \w = 2\pi\times50\,\text{Hz}$ was used to set a constraint. This is then reproduced with approximate AURIGA as the thicker dark grey line below true AURIGA. We use the mechanical energy-limited expression in Eq.~\eqref{eq:mech} to obtain this estimate, assuming an effective bar mass of $M = 1.1\times 10^3\,\text{kg}$, a length of $L = 3\,\text{m}$, a mechanical quality factor of $Q = 10^6$, and a bar temperature of $T = 0.14\,\text{K}$, corresponding to $\bar{n} \sim 10^6$ thermal phonons.

For our approximate MWB-DMR (light green), we take the parameters of the DM Radio GUT magnet, i.e. a stored EM energy of $U_{\rm in} \sim 5\times 10^9\,\text{J}$, with a typical length of $L \sim 2\,\text{m}$~\cite{DMRadio:2022jfv}. We use these parameters in the broadband case of Eq.~\eqref{eq:mech}, obtaining the light green line.  We see that our approximate MWB-DMR sensitivity matches very closely the sensitivity obtained from the noise-equivalent strain for MWB-DMR given in Ref.~\cite{Domcke:2024mfu}. For both real and approximate, we assume an integration time of $t_{\rm int} = 1\,\text{yr}$, and a bandwidth $\Delta \w = \wg$.\footnote{On the mechanical resonance, the bandwidth is $\Delta \w = \wg/Q_m$, which negates the benefit of the resonance in terms of sensitivity to $\Omega_g$, as discussed in Section~\ref{sec:mech}.}
 
For approximate MAGO 2.0 (dashed red), we use an effective volume of $1\,\text{m}^3$ and a magnetic field of $B = 0.1\,\text{T}$ for a total EM energy of $U_{\rm in} = 4\,\text{kJ}$. The intrinsic quality factor of a superconducting RF cavity such as the one used for MAGO is $Q = 10^{10}$, but for noise reasons, we have used $Q=10^5$ to estimate the sensitivity in the figure (see Ref.~\cite{Berlin:2023grv}). We use the mechanical transfer function of Eq.~\eqref{eq:Tmaster} in the limit $\wg \gg \w_m$, and assuming that $\w_1 = \w = 2\pi\,\text{GHz}$ to obtain the dashed red line in Fig.~\ref{fig:results}. We use dashes for the MAGO 2.0 curves to indicate that they are obtained using a resonant experiment. As such, we limit the total integration time for the full frequency range to $t_{\rm tot} = 1\,\text{yr}$, meaning that we set the integration time at a given frequency to $t_{\rm int} \sim 10\,\text{s}$.

The heuristic sensitivities derived with our simplified toy model detectors are evidently validated by comparison with existing and proposed detection schemes, since our goal was to provide a lower bound to the sensitivity of a given detector. Therefore, we turn to the question of examining the prospects for probing cosmogenic GWs at frequencies above kHz. In Figs.~\ref{fig:crazy_classical}. and~\ref{fig:crazy_quantum} we illustrate a general conclusion that can be drawn from our estimates on primordial GW backgrounds. Even extrapolating to ludicrous extremes existing detection schemes, achieving a sensitivity that goes beyond the BBN bound (solid orange line) appears implausible for frequencies above a MHz.\footnote{Note that our representation of the BBN bound is quite generous to a potential detector wanting to improve on it, as we are simply plotting a flat line at $\Omega_g \sim 10^{-6}$ to guide the eye, while the actual bound is on the integrated energy density as shown in Eq.~\eqref{eq:BBN}.} Additionally, reaching a MHz requires a great leap of faith in the future capabilities of mankind. 

 To be more concrete let us start by describing the estimated sensitivities in Fig.~\ref{fig:crazy_classical}. There we show a few detection concepts with quantum-limited sensitivity. In orange we plot two cross-correlated LIGO-sized interferometers limited only by shot-noise detecting a signal with
$\Delta \w= 2\pi \times \text{kHz}$ over a data taking period of one year. The power in their laser is $10^9\,\text{W}$, corresponding to $1\%$ of the US power grid (hence the label {\it NYC-laser} in the Figure).  In green we plot a bar detector modelled on ITER. For the solid (dashed) green line, we assume the ITER magnetised volume is operated as a Magnetic Weber Bar operated as a broadband (resonant with $Q = 10^6$) detector. This corresponds to a stored EM energy of $\simeq 10^{12}$~J, i.e. comparable to that stored in ITER's $11.8\,\text{T}$ magnets applied to its $8500\,\text{m}^3$ volume. The resonant setup has $Q=10^6$ and $t_{\rm in}=\text{yr}$ at each resonant frequency (i.e. more than a million years would be needed to scan the whole range plotted). 
Finally, the dotted green line shows the sensitivity of an ITER-like device ($M = 2.3\times 10^7\,\text{kg}$ ) operated as a traditional Weber bar, i.e. on a mechanical resonance with $Q_m=10^6$, with $t_{\rm in}=\text{yr}$ at each resonant frequency. We assume the device is cooled down to $T=\text{mK}$, which corresponds to $T/\w \gtrsim 10^3$. 

In spite of these extreme assumptions, none of our classical detectors at their SQL can improve over the BBN bound above 10 kHz. In Fig.~\ref{fig:crazy_quantum} we show what one can do with extreme assumptions regarding our ability to manipulate quantum states. In orange we show two cross-correlated LIGO-sized interferometers with a 1kW laser and the same assumptions on other parameters as the NYC-laser interferometer. In one case (dash-dotted line) the interferometers are operated in the squeezed vacuum limit of Eq.~\eqref{eq:lin_squeeze}, on the solid line they are operated at their Heisenberg limit, i.e. the ultimate limit on the sensitivity of any quantum experiment consistent with Heisenberg's uncertainty principle and the maximal rate of time evolution in quantum mechanics~\cite{MARGOLUS1998188}. To reach this limit, in addition to preparing and finely controlling an appropriate quantum state with all the $N_{\gamma}$ photons in the laser for the duration of the experiment one needs to keep optical losses well below $1/N_{\gamma}$~\cite{Escher:2011eff}. In green (dashed and solid lines) we show a Magnetic Weber Bar with a magnetised volume of $B=10\,\text{T}$ in a cubic metre, and $Q=10^6$. We assume this experiment is purely signal limited, i.e. we imagined reducing the noise in the quadrature that we are measuring to well below a single photon during the whole lifetime of the experiment. 

We present the curves in Figs.~\ref{fig:crazy_classical} and~\ref{fig:crazy_quantum} to show just how difficult it is to probe cosmogenic GWs above kHz. We do not think that these are realistic experimental concepts with current technology. We do not know how to build or operate them, or even if they will ever be technologically feasible. However, even assuming they are, their sensitivity gets rapidly much worse than the BBN bound above $f_g \simeq$~MHz, proving our point on the difficulty of detecting primordial high frequency backgrounds. 

There are interesting experimental proposals and existing detectors that operate at the frequencies that we are focusing on~\cite{Willke_2002, Carney:2024zzk, PhysRevLett.101.101101, Schmieden:2023fzn, Navarro:2023eii}, but that we did not list explicitly in this work. Our heuristics apply also to them and they do not change our qualitative conclusion on primordial GW backgrounds. 
Finally, here we only considered generic mechanical and EM couplings of GWs and existing detector concepts. If there is a way of engineering an enormous transfer function in a way that we have not considered (but note that the ones in our plots can be as large as $10^{12}$), it would greatly aid the effort to search for primordial stochastic GWs at high frequencies.

The detection of GWs by LIGO is a fantastic technological achievement built on decades of research and development. As can be seen in Fig.~\ref{fig:results}, the result is a detector which operates at its optimum classical level (i.e. the SQL) and even slightly beyond~\cite{PhysRevX.13.041021,membersoftheLIGOScientific:2024elc}. 
However, given a finite energy in the detector and integration time, it seems that the path forward is through quantum resources. If we can manipulate large collections of quanta to reduce noise, even a relatively modest amount of energy can improve upon existing proposals. At the time of writing it is hard to judge if it will be more feasible to appropriately harness the necessary quanta, or to simply increase the energy in a classical detector by orders of magnitude. 

\acknowledgments{We thank Asher Berlin, Valerie Domcke, G\'eraldine Servant, Nick Rodd, and Liantao Wang for valuable discussions. The work of SARE was supported by SNF Ambizione grant PZ00P2\_193322, \textit{New frontiers from sub-eV to super-TeV}. The work of RTD was partially supported by  ANR grant ANR-23-CE31-0024 {\it EUHiggs}.}

\newpage

\appendix
\interfootnotelinepenalty=10000 

\section{Power Spectral Densities}\label{app:PSD}
In this section we buttress our arguments in the main body regarding the presence or absence of time-domain auto-correlation functions of background and signal fields by writing these functions explicitly in the frequency domain. 
Our Fourier transform convention is
\be
f(t) = \int_{-\infty}^{\infty} e^{i \omega t} f( \omega) \frac{d\omega}{2\pi}\, , \quad f(\omega) = \int_{-\infty}^{\infty} e^{-i \omega t} f(t) dt \, ,\nn
\ee
In what follows we often suppress the limits of integration when they are $\pm \infty$ to improve readability.
The linear-in-strain auto-correlation function is given by
\begin{align}
    \langle E_0(t) E^*_h(t) \rangle = \frac{1}{2T}\int_{-T}^T dt\, E_0(t) E^*_h(t) = \frac{1}{2T(2\pi)^2} \int dt d\w_1 d\w_2 \,\langle E_0(\w_1) E_h^*(\w_2) \rangle e^{i(\w_1-\w_2)t} \ ,
\end{align}
where, to make contact with the draft we have included on the right-hand side an ensemble average in frequency space $\langle \cdot \rangle$, which is relevant to stochastic signals. In the following we omit it for brevity. 

We always take $T$ to be much larger than the typical period of the GWs and of $E_0$, so we can approximate the time integral over the exponential factor as a delta-function centered at $\w_1 - \w_2$. We can then easily perform the $\w_2$ integral to find
\begin{align}
    \langle E_0(t) E^*_h(t) \rangle &\propto \frac{1}{2\pi} \int d\w_1 \, \langle E_0(\w_1) E_h^*(\w_1) \rangle \ ,
    \label{eq:E0Eh}
\end{align}
i.e. that the linear-in-strain auto-correlation function is only non-zero if $E_0$ and $E_h$ both have support at a common frequency, as claimed in the main text. We could have found this result also by decomposing $E_h(t) = E_0(t) h(t)$, whose Fourier transform is then
\begin{align}
    E_h(\w) = \frac{1}{2\pi} \int d\w_1 E_0(\w_1) h(\w-\w_1) = \frac{1}{2\pi} \int d\w_1 E_0(\w-\w_1) h(\w_1) \ .
    \label{eq:EhFourier}
\end{align}
Then, evaluating the 3-point auto-correlation function gives
\begin{align}
\langle E_0(t) E_0^*(t) h^*(t) \rangle &= \frac{1}{2T}\int_{-T}^T dt\, E_0(t) E^*_0(t)h^*(t)\\
    \nonumber &= \frac{1}{2T(2\pi)^3}\int_{-T}^T dt \int d\w_1 d\w_2 d\w_3 \,E_0(\w_1) E_0^*(\w_2) h^*(\w_3) e^{i(\w_1-\w_2-\w_3)t} \\
    \nonumber&\simeq \frac{1}{2T(2\pi)^2} \int d\w_1 d\w_2 d\w_3\,E_0(\w_1) E_0^*(\w_2) h^*(\w_3) \delta(\w_1-\w_2-\w_3) \\
    \nonumber&\simeq \frac{1}{2T(2\pi)^2} \int d\w_1 d\w_2\,E_0(\w_1) E_0^*(\w_2) h^*(\w_1-\w_2) \\
    & = \frac{1}{(2\pi) 2 T} \int d\w_1\,E_0(\w_1) E_h^*(\w_1) \ ,
\end{align}
where in the last line we have used Eq.~\eqref{eq:EhFourier} to perform the $\w_2$ integral.
Similarly, we can write the quadratic-in-strain auto-correlation function as 
\begin{align}
\label{eq:PSD}
    \langle E_h(t) E^*_h(t) \rangle &= \frac{1}{(2\pi)^{2}} \int d\w_1 d\w_2\,S_{E_0}(\w_1) S_h(\w_1-\w_2) \ .
\end{align}
We see that instead of the frequency-space $h$, we have the strain Power Spectral Density (PSD) $S_h(\w)$ appearing. We define PSDs of scalar quantities $O$ that depend only on frequency as
\be
\langle O(\w) O^*(\w^\prime) \rangle = 2\pi\delta(\w-\w^\prime) S_O(\w)\, .
\ee
Our conventions preserve the usual relation between the time average of a function and its PSD
\begin{align}
\label{eq:PSD}
    \langle |f(t)|^2 \rangle &= \frac{1}{2\pi} \int d\w \, S_f(\w)  \ .
\end{align}

The conclusion we draw from Eq.~\eqref{eq:PSD} is that the integral is non-zero under more general circumstances than the linear-in-strain expression of Eq.~\eqref{eq:E0Eh} above. In particular, it can be non-zero even for a GW oscillating at $\w_1-\w_2 = \wg$ with $S_{E_0}(\w_1)$ only having support at $\w_1 = 0$. This is the scenario typically studied in resonant detectors such as EM cavities or Weber bars.

\section{Strain and Energy Density}\label{app:energy}
In this section we give a definition of the energy density $\rho_g$ of a gravitational wave and comment on its relation to the strain measured by a detector. The energy density can be defined as
\be
\rho_g = \frac{1}{32\pi G_N} \langle \dot h_{ab}(t) \dot h^{ab}(t)\rangle\, , 
\ee
where $\langle \cdot \rangle$ is a time average taken over many oscillations of the wave, just as in the previous section, and $h_{ab}$ is the spacial part of the small fluctuations of the metric around a flat Minkowski background. We can write it in terms of a function $\tilde h_A$ in frequency space as
\be
h_{ab}(t)=\sum_{A=+,\times}\int_{-\infty}^{+\infty}\frac{d\w}{2\pi} e^{-i \w t} \int d\Omega_2 \; \tilde h_A(\w, \Omega_2)e_{ab}^A(\Omega_2)\, .
\ee
Here $\Omega_2$ is a unit vector representing the direction of propagation of the wave and $e_{ab}^A$ are polarization tensors. They can be written in terms of the vectors $m, n$ that are obtained by finding two vectors orthogonal to $\Omega_2$ and each other
\be
e_{ab}^+(\Omega_2)= m_a m_b - n_a n_b\, , \nn \\
e_{ab}^\times(\Omega_2)= m_a n_b + n_a m_b\, .
\ee
Most of our definitions in the above equations follow the conventions of~\cite{Maggiore:1999vm}. The one important difference is that we use PSDs defined in the whole $\pm \infty$ range for all quantities, including noise. So our PSD for $\tilde h_A$ is
\be
\langle \tilde h_A(\w, \Omega_2) \tilde h^*_{A^\prime} (\w^\prime, \Omega_2^\prime) \rangle = \delta_{AA^\prime} \frac{\delta^2(\Omega_2 \Omega_2^\prime)}{4\pi}\delta(\w-\w^\prime) S_h(\w)\, .
\ee
It is customary to characterize primordial backgrounds of GWs in terms of their relative energy density, defined as
\be
\quad \Omega_g(\w) \equiv \frac{8\pi G_N}{3 H_0^2}\frac{d \rho_g(\w)}{d\log \w}\, .
\ee
Combining the two previous equations, we can write $\Omega_g(\w)$ in terms of strain,
\be
\langle h_{ab}(t) h^{ab}(t)\rangle =  6 H_0^2\int_{-\infty}^{+\infty} d \log \w \frac{\Omega_g(\w)}{\w^2}\, .
\ee
In the main text we neglect the details of the detector geometry and GW polarization and we make the rough approximation
\be
\langle h_{ab}(t) h^{ab}(t)\rangle \simeq h^2\, ,
\ee
where for us $h$ is the dimensionless strain measured by the detector.
We report here the relevant conceptual steps to compute the $\mathcal{O}(1)$ numbers neglected in the main text, following~\cite{Maggiore:1999vm}. 
We have to relate $\langle h_{ab}(t) h^{ab}(t)\rangle $ to a scalar quantity $s(t)$ representing the signal measured by the detector. It is common to introduce a detector tensor $D^{ab}$ that encapsulates its geometry and interaction with the GW
\be
s(t)=D^{ab}h_{ab}(t)\, , \label{eq:sig}
\ee
and detector pattern functions
\be
F^A(\hat \Omega)\equiv D^{ab} e_{ab}^A(\hat \Omega)\, ,
\ee
where $A+=, \times$ gives the polarization of the GW. Then, for a single detector 
\be
\langle s(t)^2\rangle = \frac{F}{2}\langle h_{ab}(t) h^{ab}(t)\rangle  , 
\ee
where
\be
F\equiv \int \frac{d \Omega_2}{4\pi}\sum_{A=+, \times} F^A(\Omega_2, \psi) F^A(\Omega_2, \psi)\, . 
\ee
values of $F$ for different detectors can be found in~\cite{Maggiore:1999vm}, and they are $\mathcal{O}(1/2)$.

If we take two interferometers and process their signals using optimal filtering, one can show that the optimal SNR is~\cite{Maggiore:1999vm}
\be
{\rm SNR}_{\rm int} = \left[\frac{8}{50}t_{\rm int}\int_{-\infty}^{+\infty} \frac{d\w}{2\pi} \gamma^2(\w)\frac{S_h^2(\w)}{S_n^2(\w)}\right]^{1/4}\, ,
\label{eq:OF}
\ee
and the detectors' details enter via the overlap function
\be
\gamma(\w)&\equiv & \frac{1}{F_{12}}\int \frac{d \Omega_2}{4\pi}\sum_{A=+, \times} F^A_1(\Omega_2, \psi) F^A_2(\Omega_2, \psi) e^{i\w \hat \Omega \cdot \Delta \vec x}\, , \nn \\
F_{12}&\equiv & \int \frac{d \Omega_2}{4\pi}\sum_{A=+, \times} F^A_1(\Omega_2, \psi) F^A_2(\Omega_2, \psi)\, ,
\ee
where $\Delta \vec x$ is the separation of the two detectors and $F_{12}$ is computed by taking the two detectors perfectly aligned. We again refer to~\cite{Maggiore:1999vm} for the calculation of $F_{1,2}^A$. The choice of SNR in Eq.~\eqref{eq:OF} is made to match numerically the single-detector case, i.e. the same value of ${\rm SNR}$ translate to the same sensitivity to $S_h$. In the main text we take $\gamma(\w)=1$ when estimating the sensitivity of cross-correlated detectors. In the next Section we are going to be more precise on this point and the relation between the ${\rm SNR}$ and the test statistic used to make a precise probabilistic statement on the signal.

\section{Test Statistic for GW Signals}\label{app:stat}
In the main text we have derived our smallest detectable strains by using setting to 1 the SNR in Eq.~\eqref{eq:SNR}. Here we show why this was justified by giving a more careful statistic treatment of the problem.

A primordial stochastic background of GWs can be modeled as a gaussian, stationary, isotropic and unpolarized random process with zero mean (see for instance Chapter 7 of~\cite{Maggiore_2007}). In this work we assume that the noise is also gaussian and stationary.  Therefore our detector measures a data stream $d(t)=h(t)+n(t)$ which is given by the sum of two normal random variables. Other sources of noise of course exist, as for instance a sudden earthquake, but they are eliminated via specific experimental procedures and do not enter in our discussion of the statistical sensitivity of the experiment.

Given these assumptions we know how to write the likelihood because the sum of two normally distributed variables $h,n$ follows a normal distribution with $\mu = \mu_h + \mu_n$ and $\sigma^2=\sigma_h^2+\sigma_n^2$, as can easily be proven using their characteristic functions $\varphi(h) = E[e^{i t h}]$.  Therefore we have for the likelihood
\be
L(d | h+n ; \theta) = \prod_{i=1}^{N} \frac{e^{- \frac{|d(\w_i)|^2}{2(S_h(\w_i)+S_n(\w_i))}}}{\sqrt{2\pi (S_h(\w_i)+S_n(\w_i))}}P(\theta)\, .
\ee
where $\theta$ are nuisance parameters that might be needed to describe signal and noise. We left them implicit in $S_{h,n}$ to improve readability. $N=\Delta \w t_{\rm int}/2\pi$ is the number of frequency bins that we have access to in the experiment.

We would like to set an upper bound on the strength of the signal $h$ that enters the likelihood through the strain PSD $S_h \propto h^2$. We are going to imagine that the data contain only noise and look for the median expected 95\% C.L. exclusion on $h$. We can do it by introducing the test statistic
\be
t_h = - 2 \log \frac{L(d | h+n ; \hat \theta_h)}{L(d | \hat h+n ; \hat \theta)}\Theta(h-\hat h)\, .
\ee
In the above expression $\hat h$ is the best fit value of $h$ in the dataset $d$ and $\hat \theta$ the best fit value of the nuisance parameters. At the numerator we fix the signal strength to be $h$ and $\hat \theta_h$ are the maximum likelihood estimators for the nuisance parameters subject to this constraint. When $\hat h$ fluctuates below zero we need to use a slightly different test statistic~\cite{Cowan:2010js}, but the difference is negligible for our discussion. For simplicity in the following we ignore nuisance parameters and imagine to know exactly signal and background. Including them broadens the test statistic distribution, weakening the exclusion, but does not change qualitatively our conclusions. This choice is conservative as it overestimates the actual sensitivity of an experiment to the signal and our goal is to show that for certain frequencies primordial backgrounds are out of reach no matter how optimistic we are on the detector setup.

\begin{figure}[!t]
\begin{center}
\includegraphics[width=0.6\textwidth]{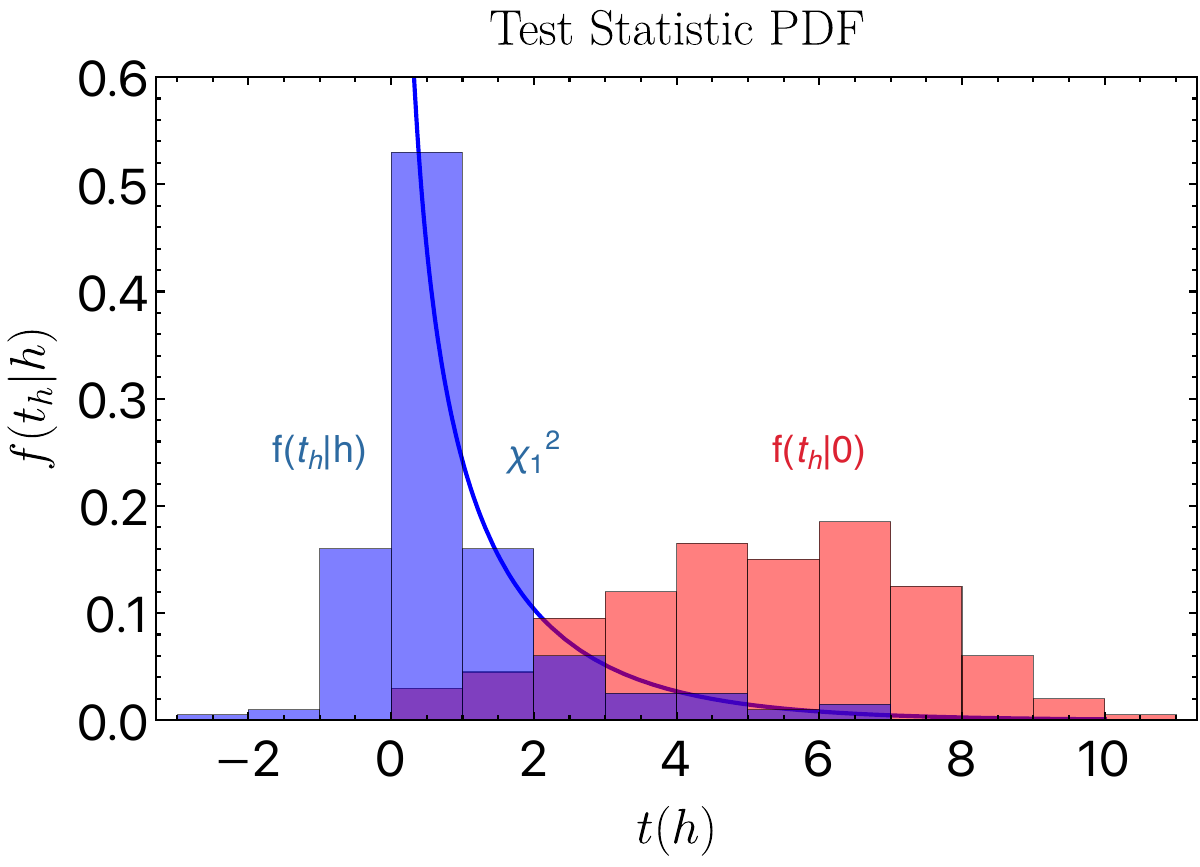}
\caption{PDF of the test statistic $t_h$ for noise-only datasets (red) and data distributed according to the signal hypothesis (blue).}
\label{fig:tpdf}
\end{center}
\end{figure}

In Fig.~\ref{fig:tpdf} we show in blue the pdf $f(t_h|h)$ of $t_h$ for data generated under the signal hypothesis $h$ that we are testing, and in red for noise-only datasets $f(t_h|0)$. The two PDFs are computed for $N=10$ and $S_h=S_n=1$, constant over $\Delta \w$. As expected from Wilks' and Wald's theorems~\cite{d543aecb-cd73-36d5-9101-f08a74f8e8c6, wald1943tests}, $f(t_h|h)$ follows a $\chi^2$ distribution with one degree of freedom, modulo some fluctuations that disappear in the asymptotic limit (i.e. for a large number of events per bin). 

Setting an upper bound on the signal strength $h$ is equivalent to rejecting the hypothesis that the test statistics follows the distribution $f(t_h|h)$, i.e. we want to find $h$ such that $t_h({\rm data})> t_{\rm cut}$, where
\be
\int_{t_{\rm cut}}^\infty d t f(t|h) = \alpha = 0.05\, .
\ee
Since we see from Fig.~\ref{fig:tpdf} that large values of $t_h$ favor the noise-only hypothesis and disfavor the signal hypothesis.
If we call $F$ the cumulative distribution of $t_h$, i.e.
\be
F(t_{\rm cut}|h)=P(t_h \leq t_{\rm cut}|h)
\ee
Then we want $F(t_{\rm cut}|h)=1-\alpha$. In the asymptotic limit $f(t_h|h) = \chi^2_1(t_h)$ and $F(t|h) = \Phi(\sqrt{t})$, where $\Phi$ is the cumulative distribution of a normal distribution. Then
\be
t_{\rm cut} = (\Phi^{-1}(1-\alpha))^2 \simeq 2.7\, .
\ee
In particular we are interested in the median expected exclusion in the presence of only known processes, so we assume that the data contain only noise and ask what $h$ we can exclude in this case. To get an analytical estimate we evaluate $t_h$ on the Asimov dataset~\cite{Cowan:2010js} $|d_i|^2 = S_n(\w_i)$ and solve
\be
t_h(d_{\rm Asimov}) = t_{\rm cut} \simeq 2.7
\label{eq:tcut}
\ee
for $h$. In the limit $N\gg 1$, i.e. $S_h \ll S_n$, and assuming $S_h$ and $S_n$ constant in the frequency interval of interest we get
\be
t_h(d_{\rm Asimov}) \simeq \frac{t_{\rm int}\Delta \w}{2\pi} \frac{S_h^2}{S_n^2}\, .
\label{eq:SNRprimordial1}
\ee
In the continuum limit and for $S_h$ and $S_n$ varying over $\Delta \w$, the median expected excluded signal can be found using the SNR
\be
t_h(d_{\rm Asimov}) \to t_{\rm int} \int \frac{d\w}{2\pi} \frac{S_h^2(\w)}{S_n^2(\w)}\, .
\label{eq:SNRprimordial2}
\ee
This reproduces the $h_{\rm min} \sim 1/t_{\rm int}^{1/4}$ scaling expected for a stochastic background of GWs. We have verified this scaling numerically, without making any assumption on the relative size of $S_h$ and $S_n$.

As a sanity check we can apply the same technique to a periodic signal that we model as a gaussian process with non-zero mean. In this case we obtain
\be
t_h^{\rm periodic}(d_{\rm Asimov}) \simeq t_{\rm int} \int \frac{d\w}{2\pi}\frac{S_h(\w)}{S_n(\w)} \gtrsim 2.7\, , \label{eq:meanSNR}
\ee
assuming $S_h$ dominated by the $\langle h(t)^2\rangle$ term which is zero for a primordial background. We reproduce the $h\sim 1/t_{\rm int}^{1/2}$ scaling that one expects from a matched filtering analysis~\cite{Maggiore_2007}. To obtain $h_{\rm min}$ in the main text we set 
\be
{\rm SNR} \equiv\sqrt{t_h(d_{\rm Asimov})} \simeq \left(t_{\rm int}\int \frac{d \w}{2\pi} \frac{S_h^2(\w)}{S_n^2(\w)}\right)^{1/2} \simeq 1 \label{eq:SNR1}
\ee
this makes contact with the expressions of the SNR that one can find in the literature and is a better approximation than setting $t_h(d_{\rm Asimov}) \simeq 1$, as one can see from Eq.~\eqref{eq:tcut}. In the main text we use Eq.~\eqref{eq:SNR1} to derive our minimal detectable strains.

\section{Coupling of Gravitational Waves to Mechanical and Electromagnetic Modes of the Detector}\label{app:EOM}
In this Section we review known results on the coupling of GWs to matter following the notation in~\cite{Berlin:2021txa, Berlin:2023grv}. This makes all the assumption that went into the transfer functions in Eq.~\eqref{eq:Tmaster} completely explicit. We first expand the electric field in the detector into normal modes
\be
\Evec(t, \xv)=\sum_n e_n(t) \Evec_n(\xv)\, ,
\ee
then we focus on the two modes $e_1$ which is readout and will contain information about the GW signal and $e_0$ that we have prepared in the laboratory with some electromagnetic energy before the arrival of the GW. 

\subsection{Mechanical Coupling}
A small deformation of the detector $V\to V+\Delta V$ can mix the unperturbed modes and act as a source for $e_1$ in the presence of energy in $e_0$. In the limit $\Delta V/V = \mathcal{O}(h) \ll 1$ we can write the wave equation for the signal mode $e_1$ as
\be
\left(\partial_t^2+\frac{\w_1}{Q}\partial_t+\w_1^2\right)e_1(t) = - \w_1^2 e_0(t)\frac{\int_{\Delta V}d^3 x \left(\E_0\cdot \E_1-(\w_0/\w_1) \B_0 \cdot \B^*_1\right)}{\int_{V_0}d^3 x |\E_1|^2}+\mathcal{O}(h^2)\, .
\ee
We can now relate $\Delta V$ to the effect of the GW. Without loss of generality we rewrite the integral over the volume deformation in terms of a displacement vector $\U$,
\be
\int_{\Delta V}d^3 x = \int_{S_0} d \A \cdot \U\, .
\ee
Then we decompose $\U$ in mechanical modes of the detector,
\be
\U(t, \xv)=\sum_\alpha u_\alpha(t) \U_\alpha(\xv)\, , \quad \sum_\alpha\int_{V_0} d^3 x \, \rho(\xv) |\U_\alpha|^2 = M\, ,
\ee
where $\rho, M$ are the density and mass of the detector, respectively. This allows us to rewrite the wave equation as 
\be
\left(\partial_t^2+\frac{\w_1}{Q}\partial_t+\w_1^2\right)e_1(t) &=& - \w_1^2 e_0(t) \sum_\alpha u_\alpha(t) C^\alpha+\mathcal{O}(h^2)\, , \\
C^\alpha&\equiv& \frac{\int_{S_0}d\A \cdot \U_\alpha \left(\E_0\cdot \E_1-(\w_0/\w_1) \B_0 \cdot \B^*_1\right)}{\int_{V_0}d^3 x |\E_1|^2}\, .
\ee
We can now write the equation of motion for $u_\alpha$ with the GW as a source. From classical, non–relativistic, linear elasticity theory we have
\be
\left(\partial_t^2+\frac{\w_\alpha}{Q}\partial_t+\w_\alpha^2\right)u_\alpha(t) =\frac{F_\alpha}{M_{\rm cav}}\, .\label{eq:mech2}
\ee
In the above equation we introduced $F_\alpha$ which is the generalized force projected onto the appropriate mechanical mode. We can define it from the force density $\mathbf{f}$ as
\be
F_\alpha(t)=\int_{V_0} d^3 x \;\mathbf{f}(t,\xv)\cdot \U_\alpha(\xv)\, .
\ee
The force density from the GW in the PDF is obtained from the equation of geodesic deviation as in Ch. 1 of~\cite{Maggiore},
\be
f_i^{\rm GW}(t,\xv)=R_{i0j0}(t,\xv) x^j \rho(\xv)\, .
\ee
In the main text (Eq.~\eqref{eq:EOM}) we took the two coupled equations that we just derived
\be
\left(\partial_t^2+\frac{\w_1}{Q}\partial_t+\w_1^2\right)e_1(t) &=& - \w_1^2 e_0(t) u_m(t) C^m\, , \\
\left(\partial_t^2+\frac{\w_m}{Q_m}\partial_t+\w_m^2\right)u_m(t) &=&\frac{F_m}{M_{\rm cav}}\, ,
\ee
assumed that one mode $\alpha=m$ predominantly couples to the gravitational wave and replaced $C^m$ and $R_{i0j0}x^j$ with approximate estimates that saturate dimensional analysis $C^m \simeq L$, $R_{i0j0}x^j \simeq \w_g^2 L h^{TT}_{ij}$ for a monochromatic wave (as we always considered signals with $\Delta \w \lesssim \w_g$). Note that these values for $C^m$ and $R_{i0j0}x^j$ can be achieved in existing detectors and some examples are given in~\cite{Berlin:2021txa, Berlin:2023grv} and by the detectors that approximately saturate our estimates in Fig.~\ref{fig:results}. Note also that the Riemann tensor in the previous Eqs. is evaluated at the origin of our coordinate system (the center of mass of the detector) and it is invariant under coordinate transformations at $\mathcal{O}(h)$. So we can evaluate it in terms of the metric in TT gauge, which for a wave propagating along the $z$ direction reads, $h_{00}^{\rm TT}=h_{0i}^{\rm TT}=0$, $h_{xx}^{\rm TT}=h_+ e^{i \w_g t}$, $h_{xy}^{\rm TT}=h_\times e^{i \w_g t}$, $h_{yy}^{\rm TT}=-h_+ e^{i \w_g t}$.

A point to keep in mind is that $V$ is the fraction of the volume of the detector that supports mechanical modes, so for example the volume of LIGO's mirrors or the volume of the walls of a resonating cavity. In this paper we never consider the limit for which the wavelength of the GW becomes smaller than $V^{1/3}$. The scale $L$ that we use in the main text is associated to the size of the whole detector and it can be much larger than $V^{1/3}$, as is the case for LVK and resonant cavities, or it can be comparable as for Weber bars.

\subsection{Electromagnetic Coupling}
The interaction of the GW with the EM field can be parametrized in terms of an effective current~\cite{Berlin:2021txa}
\be
\label{eq:jeff}
\jeff^\mu \equiv 
\partial_\nu \left( \frac{1}{2} \, h \, F^{\mu \nu} + h^\nu_{~ \alpha} \, F^{\alpha \mu} - h^\mu_{~ \alpha} \, F^{\alpha \nu} \right)
~,
\ee
that we can write explicitly using the metric in the PDF~\cite{Marzlin:1994ia}. For a monochromatic GW propagating along the $z$ direction it reads
\begin{align}
\label{eq:metric}
h_{00} &= - R_{0i0j} \, x^i \, x^j \times 2 \left[ - \frac{i}{k_g z} + \frac{1 - e^{- i k_g z}}{(k_g z)^2} \right]\, ,
\nl
h_{ij} &= - \frac{1}{3} \, R_{ikjl} \, x^k \, x^l \times 6 \left[ - \frac{1 + e^{- i k_g z}}{(k_g z)^2} - 2 i \,  \frac{1 - e^{- i k_g z}}{(k_g z)^3}  \right]\, ,
\nl
h_{0i} &= - \frac{2}{3} \, R_{0jik} \, x^j \, x^k \times 3 \left[ - \frac{i}{2 \, k_g z} - \frac{e^{- i k_g z}}{(k_g z)^2} - i \,  \frac{1 - e^{- i k_g z}}{(k_g z)^3}  \right]
~.
\end{align}
In the above expression the Riemann tensor is evaluated at the center of mass of the detector, which is the origin of the PDF. Recalling that the Riemann tensor is invariant under coordinate transformations at $\mathcal{O}(h)$ we can compute it in terms of the metric in TT gauge, that we give here for a wave propagating along the $z$ direction, $h_{00}^{\rm TT}=h_{0i}^{\rm TT}=0$, $h_{xx}^{\rm TT}=h_+ e^{i \w_g t}$, $h_{xy}^{\rm TT}=h_\times e^{i \w_g t}$, $h_{yy}^{\rm TT}=-h_+ e^{i \w_g t}$. This explains the apperance of $h^{\rm TT}$ in Eq.~\eqref{eq:EOM}, as already discussed in the previous Section.
In Eq.~\eqref{eq:metric} we show explicitly the difference between $k_g$ and $\w_g$ because most detectors can observe many oscillations of the GW in the time domain while being small compared to $1/k_g$ and observing an approximately constant spatial profile of the wave. In the main text we set $c=1$ and $k_g =\w_g$, but the reader interested in reproducing our transfer functions might find useful to keep them distinct.

Starting from Eq.~\eqref{eq:jeff} we can write the wave equation for our signal mode $e_1$,
\be
\Big(\partial_t^2 +\frac{\w_1}{Q} \, \partial_t + \w_1^2 \Big) \, e_1(t)=
	- \, \frac{\int_{\Vcav} \hspace{-0.2cm} d^3 \xv ~ \E_1^* \cdot \partial_t \jveff(t)}{\int_{\Vcav} \hspace{-0.2cm} d^3 \xv ~ |\E_1|^2}
	~\, . \label{eq:EM}
\ee
To highlight the parametric dependence of the above equation on $\w_g$ and the size of the detector $L$ we can assume that the effective current is given mostly by the pump mode $e_0$ and the GW signal is not too broad $\Delta \w \lesssim \w_g$. Then if we define
\be
\jveff \equiv \w_g (1+\w_g L +\w_0 L)\min[1, \w_g L]\overline{ \jveff}\, ,
\ee
$\overline{\jveff}$ is a $\mathcal{O}(1)$ number in both the long ($k_g \to 0$) and short wavelength limit of the GW. To derive Eq.~\eqref{eq:EOM} we took Eq.~\eqref{eq:EM} and set the overlap factor
\be
\frac{\int_{\Vcav} \hspace{-0.2cm} d^3 \xv ~ \E_1^* \cdot \overline{\jveff}}{\int_{\Vcav} \hspace{-0.2cm} d^3 \xv ~ |\E_1|^2} \simeq 1\, ,
\ee
which can be achieved by an appropriate choice of the detector geometry~\cite{Berlin:2021txa, Berlin:2023grv}.

\bibliography{biblio}

\providecommand{\href}[2]{#2}\begingroup\raggedright\begin{thebibliography}{100}

\bibitem{Giovannetti:2011chh}
V.~Giovannetti, S.~Lloyd and L.~Maccone, \emph{{Advances in quantum
  metrology}}, \href{https://doi.org/10.1038/nphoton.2011.35}{\emph{Nature
  Photon.} {\bfseries 5} (2011) 222}
  [\href{https://arxiv.org/abs/1102.2318}{{\ttfamily 1102.2318}}].

\bibitem{RevModPhys.90.035005}
L.~Pezz\`e, A.~Smerzi, M.K.~Oberthaler, R.~Schmied and P.~Treutlein,
  \emph{Quantum metrology with nonclassical states of atomic ensembles},
  \href{https://doi.org/10.1103/RevModPhys.90.035005}{\emph{Rev. Mod. Phys.}
  {\bfseries 90} (2018) 035005}.

\bibitem{Brito:2015oca}
R.~Brito, V.~Cardoso and P.~Pani, \emph{{Superradiance}: {New Frontiers in
  Black Hole Physics}},
  \href{https://doi.org/10.1007/978-3-319-19000-6}{\emph{Lect. Notes Phys.}
  {\bfseries 906} (2015) pp.1}
  [\href{https://arxiv.org/abs/1501.06570}{{\ttfamily 1501.06570}}].

\bibitem{Arvanitaki:2010sy}
A.~Arvanitaki and S.~Dubovsky, \emph{{Exploring the String Axiverse with
  Precision Black Hole Physics}},
  \href{https://doi.org/10.1103/PhysRevD.83.044026}{\emph{Phys. Rev. D}
  {\bfseries 83} (2011) 044026}
  [\href{https://arxiv.org/abs/1004.3558}{{\ttfamily 1004.3558}}].

\bibitem{Arvanitaki:2014wva}
A.~Arvanitaki, M.~Baryakhtar and X.~Huang, \emph{{Discovering the QCD Axion
  with Black Holes and Gravitational Waves}},
  \href{https://doi.org/10.1103/PhysRevD.91.084011}{\emph{Phys. Rev. D}
  {\bfseries 91} (2015) 084011}
  [\href{https://arxiv.org/abs/1411.2263}{{\ttfamily 1411.2263}}].

\bibitem{Casalderrey-Solana:2022rrn}
J.~Casalderrey-Solana, D.~Mateos and M.~Sanchez-Garitaonandia,
  \emph{{Mega-Hertz Gravitational Waves from Neutron Star Mergers}},
  \href{https://arxiv.org/abs/2210.03171}{{\ttfamily 2210.03171}}.

\bibitem{Franciolini:2022htd}
G.~Franciolini, A.~Maharana and F.~Muia, \emph{{Hunt for light primordial black
  hole dark matter with ultrahigh-frequency gravitational waves}},
  \href{https://doi.org/10.1103/PhysRevD.106.103520}{\emph{Phys. Rev. D}
  {\bfseries 106} (2022) 103520}
  [\href{https://arxiv.org/abs/2205.02153}{{\ttfamily 2205.02153}}].

\bibitem{Franciolini:2022ewd}
G.~Franciolini, K.~Kritos, E.~Berti and J.~Silk, \emph{{Primordial black hole
  mergers from three-body interactions}},
  \href{https://doi.org/10.1103/PhysRevD.106.083529}{\emph{Phys. Rev. D}
  {\bfseries 106} (2022) 083529}
  [\href{https://arxiv.org/abs/2205.15340}{{\ttfamily 2205.15340}}].

\bibitem{Aggarwal:2020olq}
N.~Aggarwal et~al., \emph{{Challenges and opportunities of gravitational-wave
  searches at MHz to GHz frequencies}},
  \href{https://doi.org/10.1007/s41114-021-00032-5}{\emph{Living Rev. Rel.}
  {\bfseries 24} (2021) 4} [\href{https://arxiv.org/abs/2011.12414}{{\ttfamily
  2011.12414}}].

\bibitem{LIGOScientific:2016aoc}
{\scshape LIGO Scientific, Virgo} collaboration, \emph{{Observation of
  Gravitational Waves from a Binary Black Hole Merger}},
  \href{https://doi.org/10.1103/PhysRevLett.116.061102}{\emph{Phys. Rev. Lett.}
  {\bfseries 116} (2016) 061102}
  [\href{https://arxiv.org/abs/1602.03837}{{\ttfamily 1602.03837}}].

\bibitem{NANOGrav:2023gor}
{\scshape NANOGrav} collaboration, \emph{{The NANOGrav 15 yr Data Set: Evidence
  for a Gravitational-wave Background}},
  \href{https://doi.org/10.3847/2041-8213/acdac6}{\emph{Astrophys. J. Lett.}
  {\bfseries 951} (2023) L8}
  [\href{https://arxiv.org/abs/2306.16213}{{\ttfamily 2306.16213}}].

\bibitem{Xu:2023wog}
H.~Xu et~al., \emph{{Searching for the Nano-Hertz Stochastic Gravitational Wave
  Background with the Chinese Pulsar Timing Array Data Release I}},
  \href{https://doi.org/10.1088/1674-4527/acdfa5}{\emph{Res. Astron.
  Astrophys.} {\bfseries 23} (2023) 075024}
  [\href{https://arxiv.org/abs/2306.16216}{{\ttfamily 2306.16216}}].

\bibitem{Reardon:2023gzh}
D.J.~Reardon et~al., \emph{{Search for an Isotropic Gravitational-wave
  Background with the Parkes Pulsar Timing Array}},
  \href{https://doi.org/10.3847/2041-8213/acdd02}{\emph{Astrophys. J. Lett.}
  {\bfseries 951} (2023) L6}
  [\href{https://arxiv.org/abs/2306.16215}{{\ttfamily 2306.16215}}].

\bibitem{EPTA:2023fyk}
{\scshape EPTA, InPTA:} collaboration, \emph{{The second data release from the
  European Pulsar Timing Array - III. Search for gravitational wave signals}},
  \href{https://doi.org/10.1051/0004-6361/202346844}{\emph{Astron. Astrophys.}
  {\bfseries 678} (2023) A50}
  [\href{https://arxiv.org/abs/2306.16214}{{\ttfamily 2306.16214}}].

\bibitem{Ghiglieri:2015nfa}
J.~Ghiglieri and M.~Laine, \emph{{Gravitational wave background from Standard
  Model physics: Qualitative features}},
  \href{https://doi.org/10.1088/1475-7516/2015/07/022}{\emph{JCAP} {\bfseries
  07} (2015) 022} [\href{https://arxiv.org/abs/1504.02569}{{\ttfamily
  1504.02569}}].

\bibitem{Giovannini:2019oii}
M.~Giovannini, \emph{{Primordial backgrounds of relic gravitons}},
  \href{https://doi.org/10.1016/j.ppnp.2020.103774}{\emph{Prog. Part. Nucl.
  Phys.} {\bfseries 112} (2020) 103774}
  [\href{https://arxiv.org/abs/1912.07065}{{\ttfamily 1912.07065}}].

\bibitem{Ringwald:2020ist}
A.~Ringwald, J.~Sch\"utte-Engel and C.~Tamarit, \emph{{Gravitational Waves as a
  Big Bang Thermometer}},
  \href{https://doi.org/10.1088/1475-7516/2021/03/054}{\emph{JCAP} {\bfseries
  03} (2021) 054} [\href{https://arxiv.org/abs/2011.04731}{{\ttfamily
  2011.04731}}].

\bibitem{Ghiglieri:2022rfp}
J.~Ghiglieri, J.~Sch\"utte-Engel and E.~Speranza, \emph{{Freezing-In
  Gravitational Waves}},  \href{https://arxiv.org/abs/2211.16513}{{\ttfamily
  2211.16513}}.

\bibitem{Giovannini:2023itq}
M.~Giovannini, \emph{{Relic gravitons and high-frequency detectors}},
  \href{https://arxiv.org/abs/2303.11928}{{\ttfamily 2303.11928}}.

\bibitem{hooper2020hot}
D.~Hooper, G.~Krnjaic, J.~March-Russell, S.D.~McDermott and
  R.~Petrossian-Byrne, \emph{Hot gravitons and gravitational waves from kerr
  black holes in the early universe},  2020.

\bibitem{Arvanitaki:2012cn}
A.~Arvanitaki and A.A.~Geraci, \emph{{Detecting high-frequency gravitational
  waves with optically-levitated sensors}},
  \href{https://doi.org/10.1103/PhysRevLett.110.071105}{\emph{Phys. Rev. Lett.}
  {\bfseries 110} (2013) 071105}
  [\href{https://arxiv.org/abs/1207.5320}{{\ttfamily 1207.5320}}].

\bibitem{Janssen:2014dka}
G.~Janssen et~al., \emph{{Gravitational wave astronomy with the SKA}},
  \href{https://doi.org/10.22323/1.215.0037}{\emph{PoS} {\bfseries AASKA14}
  (2015) 037} [\href{https://arxiv.org/abs/1501.00127}{{\ttfamily
  1501.00127}}].

\bibitem{Namikawa:2019tax}
T.~Namikawa, S.~Saga, D.~Yamauchi and A.~Taruya, \emph{{CMB Constraints on the
  Stochastic Gravitational-Wave Background at Mpc scales}},
  \href{https://doi.org/10.1103/PhysRevD.100.021303}{\emph{Phys. Rev. D}
  {\bfseries 100} (2019) 021303}
  [\href{https://arxiv.org/abs/1904.02115}{{\ttfamily 1904.02115}}].

\bibitem{CMB-S4:2020lpa}
{\scshape CMB-S4} collaboration, \emph{{CMB-S4: Forecasting Constraints on
  Primordial Gravitational Waves}},
  \href{https://arxiv.org/abs/2008.12619}{{\ttfamily 2008.12619}}.

\bibitem{Badurina:2019hst}
L.~Badurina et~al., \emph{{AION: An Atom Interferometer Observatory and
  Network}}, \href{https://doi.org/10.1088/1475-7516/2020/05/011}{\emph{JCAP}
  {\bfseries 05} (2020) 011}
  [\href{https://arxiv.org/abs/1911.11755}{{\ttfamily 1911.11755}}].

\bibitem{Abe:2021ksx}
M.~Abe et~al., \emph{{Matter-wave Atomic Gradiometer Interferometric Sensor
  (MAGIS-100)}}, \href{https://doi.org/10.1088/2058-9565/abf719}{\emph{Quantum
  Sci. Technol.} {\bfseries 6} (2021) 044003}
  [\href{https://arxiv.org/abs/2104.02835}{{\ttfamily 2104.02835}}].

\bibitem{AEDGE:2019nxb}
{\scshape AEDGE} collaboration, \emph{{AEDGE: Atomic Experiment for Dark Matter
  and Gravity Exploration in Space}},
  \href{https://doi.org/10.1140/epjqt/s40507-020-0080-0}{\emph{EPJ Quant.
  Technol.} {\bfseries 7} (2020) 6}
  [\href{https://arxiv.org/abs/1908.00802}{{\ttfamily 1908.00802}}].

\bibitem{Hild:2010id}
S.~Hild et~al., \emph{{Sensitivity Studies for Third-Generation Gravitational
  Wave Observatories}},
  \href{https://doi.org/10.1088/0264-9381/28/9/094013}{\emph{Class. Quant.
  Grav.} {\bfseries 28} (2011) 094013}
  [\href{https://arxiv.org/abs/1012.0908}{{\ttfamily 1012.0908}}].

\bibitem{Punturo:2010zz}
M.~Punturo et~al., \emph{{The Einstein Telescope: A third-generation
  gravitational wave observatory}},
  \href{https://doi.org/10.1088/0264-9381/27/19/194002}{\emph{Class. Quant.
  Grav.} {\bfseries 27} (2010) 194002}.

\bibitem{LIGOScientific:2016wof}
{\scshape LIGO Scientific} collaboration, \emph{{Exploring the Sensitivity of
  Next Generation Gravitational Wave Detectors}},
  \href{https://doi.org/10.1088/1361-6382/aa51f4}{\emph{Class. Quant. Grav.}
  {\bfseries 34} (2017) 044001}
  [\href{https://arxiv.org/abs/1607.08697}{{\ttfamily 1607.08697}}].

\bibitem{LISA:2017pwj}
{\scshape LISA} collaboration, \emph{{Laser Interferometer Space Antenna}},
  \href{https://arxiv.org/abs/1702.00786}{{\ttfamily 1702.00786}}.

\bibitem{Yagi:2011wg}
K.~Yagi and N.~Seto, \emph{{Detector configuration of DECIGO/BBO and
  identification of cosmological neutron-star binaries}},
  \href{https://doi.org/10.1103/PhysRevD.83.044011}{\emph{Phys. Rev. D}
  {\bfseries 83} (2011) 044011}
  [\href{https://arxiv.org/abs/1101.3940}{{\ttfamily 1101.3940}}].

\bibitem{Goryachev:2014yra}
M.~Goryachev and M.E.~Tobar, \emph{{Gravitational Wave Detection with High
  Frequency Phonon Trapping Acoustic Cavities}},
  \href{https://doi.org/10.1103/PhysRevD.90.102005}{\emph{Phys. Rev. D}
  {\bfseries 90} (2014) 102005}
  [\href{https://arxiv.org/abs/1410.2334}{{\ttfamily 1410.2334}}].

\bibitem{Ejlli:2019bqj}
A.~Ejlli, D.~Ejlli, A.M.~Cruise, G.~Pisano and H.~Grote, \emph{{Upper limits on
  the amplitude of ultra-high-frequency gravitational waves from graviton to
  photon conversion}},
  \href{https://doi.org/10.1140/epjc/s10052-019-7542-5}{\emph{Eur. Phys. J. C}
  {\bfseries 79} (2019) 1032}
  [\href{https://arxiv.org/abs/1908.00232}{{\ttfamily 1908.00232}}].

\bibitem{Aggarwal:2020umq}
N.~Aggarwal, G.P.~Winstone, M.~Teo, M.~Baryakhtar, S.L.~Larson, V.~Kalogera
  et~al., \emph{{Searching for New Physics with a Levitated-Sensor-Based
  Gravitational-Wave Detector}},
  \href{https://doi.org/10.1103/PhysRevLett.128.111101}{\emph{Phys. Rev. Lett.}
  {\bfseries 128} (2022) 111101}
  [\href{https://arxiv.org/abs/2010.13157}{{\ttfamily 2010.13157}}].

\bibitem{Berlin:2021txa}
A.~Berlin, D.~Blas, R.~Tito~D'Agnolo, S.A.R.~Ellis, R.~Harnik, Y.~Kahn et~al.,
  \emph{{Detecting high-frequency gravitational waves with microwave
  cavities}}, \href{https://doi.org/10.1103/PhysRevD.105.116011}{\emph{Phys.
  Rev. D} {\bfseries 105} (2022) 116011}
  [\href{https://arxiv.org/abs/2112.11465}{{\ttfamily 2112.11465}}].

\bibitem{Domcke:2022rgu}
V.~Domcke, C.~Garcia-Cely and N.L.~Rodd, \emph{{Novel Search for High-Frequency
  Gravitational Waves with Low-Mass Axion Haloscopes}},
  \href{https://doi.org/10.1103/PhysRevLett.129.041101}{\emph{Phys. Rev. Lett.}
  {\bfseries 129} (2022) 041101}
  [\href{https://arxiv.org/abs/2202.00695}{{\ttfamily 2202.00695}}].

\bibitem{Berlin:2023grv}
A.~Berlin, D.~Blas, R.~Tito~D'Agnolo, S.A.R.~Ellis, R.~Harnik, Y.~Kahn et~al.,
  \emph{{Electromagnetic cavities as mechanical bars for gravitational waves}},
  \href{https://doi.org/10.1103/PhysRevD.108.084058}{\emph{Phys. Rev. D}
  {\bfseries 108} (2023) 084058}
  [\href{https://arxiv.org/abs/2303.01518}{{\ttfamily 2303.01518}}].

\bibitem{Bringmann:2023gba}
T.~Bringmann, V.~Domcke, E.~Fuchs and J.~Kopp, \emph{{High-frequency
  gravitational wave detection via optical frequency modulation}},
  \href{https://doi.org/10.1103/PhysRevD.108.L061303}{\emph{Phys. Rev. D}
  {\bfseries 108} (2023) L061303}
  [\href{https://arxiv.org/abs/2304.10579}{{\ttfamily 2304.10579}}].

\bibitem{Domcke:2023bat}
V.~Domcke, C.~Garcia-Cely, S.M.~Lee and N.L.~Rodd, \emph{{Symmetries and
  selection rules: optimising axion haloscopes for Gravitational Wave
  searches}}, \href{https://doi.org/10.1007/JHEP03(2024)128}{\emph{JHEP}
  {\bfseries 03} (2024) 128}
  [\href{https://arxiv.org/abs/2306.03125}{{\ttfamily 2306.03125}}].

\bibitem{Tobar:2023gvp}
M.E.~Tobar, \emph{{Gravitational Wave Detection and Low-Noise Sapphire
  Oscillators}},  other thesis, 11, 2023,
  [\href{https://arxiv.org/abs/2311.16426}{{\ttfamily 2311.16426}}].

\bibitem{Kahn:2023mrj}
Y.~Kahn, J.~Sch\"utte-Engel and T.~Trickle, \emph{{Searching for high-frequency
  gravitational waves with phonons}},
  \href{https://doi.org/10.1103/PhysRevD.109.096023}{\emph{Phys. Rev. D}
  {\bfseries 109} (2024) 096023}
  [\href{https://arxiv.org/abs/2311.17147}{{\ttfamily 2311.17147}}].

\bibitem{Domcke:2024mfu}
V.~Domcke, S.A.R.~Ellis and N.L.~Rodd, \emph{{Magnets are Weber Bar
  Gravitational Wave Detectors}},
  \href{https://arxiv.org/abs/2408.01483}{{\ttfamily 2408.01483}}.

\bibitem{Carney:2024zzk}
D.~Carney, G.~Higgins, G.~Marocco and M.~Wentzel, \emph{{A Superconducting
  Levitated Detector of Gravitational Waves}},
  \href{https://arxiv.org/abs/2408.01583}{{\ttfamily 2408.01583}}.

\bibitem{Domcke:2024eti}
V.~Domcke, S.A.R.~Ellis and J.~Kopp, \emph{{Dielectric Haloscopes as
  Gravitational Wave Detectors}},
  \href{https://arxiv.org/abs/2409.06462}{{\ttfamily 2409.06462}}.

\bibitem{Capdevilla:2024cby}
R.~Capdevilla, G.B.~Gelmini, J.~Hyman, A.J.~Millar and E.~Vitagliano,
  \emph{{Gravitational Wave Detection With Plasma Haloscopes}},
  \href{https://arxiv.org/abs/2412.14450}{{\ttfamily 2412.14450}}.

\bibitem{Escher:2011eff}
B.M.~Escher, R.L.~de~Matos~Filho and L.~Davidovich, \emph{{General framework
  for estimating the ultimate precision limit in noisy quantum-enhanced
  metrology}}, \href{https://doi.org/10.1038/nphys1958}{\emph{Nature Phys.}
  {\bfseries 7} (2011) 406}.

\bibitem{clerkIntroductionQuantumNoise2010}
A.A.~Clerk, M.H.~Devoret, S.M.~Girvin, F.~Marquardt and R.J.~Schoelkopf,
  \emph{Introduction to {{Quantum Noise}}, {{Measurement}} and
  {{Amplification}}},
  \href{https://doi.org/10.1103/RevModPhys.82.1155}{\emph{Reviews of Modern
  Physics} {\bfseries 82} (2010) 1155}
  [\href{https://arxiv.org/abs/0810.4729}{{\ttfamily 0810.4729}}].

\bibitem{Aspelmeyer:2013lha}
M.~Aspelmeyer, T.J.~Kippenberg and F.~Marquardt, \emph{{Cavity Optomechanics}},
  \href{https://doi.org/10.1103/RevModPhys.86.1391}{\emph{Rev. Mod. Phys.}
  {\bfseries 86} (2014) 1391}
  [\href{https://arxiv.org/abs/1303.0733}{{\ttfamily 1303.0733}}].

\bibitem{Beckey:2023shi}
J.~Beckey, D.~Carney and G.~Marocco, \emph{{Quantum measurements in fundamental
  physics: a user's manual}},
  \href{https://arxiv.org/abs/2311.07270}{{\ttfamily 2311.07270}}.

\bibitem{Mangano:2011ar}
G.~Mangano and P.D.~Serpico, \emph{{A robust upper limit on $N_{\rm eff}$ from
  BBN, circa 2011}},
  \href{https://doi.org/10.1016/j.physletb.2011.05.075}{\emph{Phys. Lett. B}
  {\bfseries 701} (2011) 296}
  [\href{https://arxiv.org/abs/1103.1261}{{\ttfamily 1103.1261}}].

\bibitem{Cyburt:2015mya}
R.H.~Cyburt, B.D.~Fields, K.A.~Olive and T.-H.~Yeh, \emph{{Big Bang
  Nucleosynthesis: 2015}},
  \href{https://doi.org/10.1103/RevModPhys.88.015004}{\emph{Rev. Mod. Phys.}
  {\bfseries 88} (2016) 015004}
  [\href{https://arxiv.org/abs/1505.01076}{{\ttfamily 1505.01076}}].

\bibitem{Peimbert:2016bdg}
A.~Peimbert, M.~Peimbert and V.~Luridiana, \emph{{The primordial helium
  abundance and the number of neutrino families}}, {\emph{Rev. Mex. Astron.
  Astrofis.} {\bfseries 52} (2016) 419}
  [\href{https://arxiv.org/abs/1608.02062}{{\ttfamily 1608.02062}}].

\bibitem{Planck:2018jri}
{\scshape Planck} collaboration, \emph{{Planck 2018 results. X. Constraints on
  inflation}}, \href{https://doi.org/10.1051/0004-6361/201833887}{\emph{Astron.
  Astrophys.} {\bfseries 641} (2020) A10}
  [\href{https://arxiv.org/abs/1807.06211}{{\ttfamily 1807.06211}}].

\bibitem{US}
R.T.~D'Agnolo and S.~Ellis, ``Quantum (and classical) detection of
  gravitational waves, \textit{In preparation}.''.

\bibitem{Vermeulen:2024vgl}
S.M.~Vermeulen et~al., \emph{{Photon Counting Interferometry to Detect
  Geontropic Space-Time Fluctuations with GQuEST}},
  \href{https://arxiv.org/abs/2404.07524}{{\ttfamily 2404.07524}}.

\bibitem{McCuller:2022hum}
L.~McCuller, \emph{{Single-Photon Signal Sideband Detection for High-Power
  Michelson Interferometers}},
  \href{https://arxiv.org/abs/2211.04016}{{\ttfamily 2211.04016}}.

\bibitem{Manasse:1963zz}
F.K.~Manasse and C.W.~Misner, \emph{{Fermi Normal Coordinates and Some Basic
  Concepts in Differential Geometry}},
  \href{https://doi.org/10.1063/1.1724316}{\emph{J. Math. Phys.} {\bfseries 4}
  (1963) 735}.

\bibitem{Misner1973}
C.W.~Misner, K.S.~Thorne and J.A.~Wheeler, \emph{{Gravitation}}, W. H. Freeman,
  San Francisco (1973).

\bibitem{Maggiore}
M.~Maggiore, \emph{{Gravitational Waves. Vol. 1: Theory and Experiments}},
  Oxford Master Series in Physics, Oxford University Press (2007).

\bibitem{Maggiore_2007}
M.~Maggiore, \emph{Gravitational waves}, Oxford University Press (2007).

\bibitem{Shi:2022wpf}
H.~Shi and Q.~Zhuang, \emph{{Ultimate precision limit of noise sensing and dark
  matter search}}, \href{https://doi.org/10.1038/s41534-023-00693-w}{\emph{npj
  Quantum Inf.} {\bfseries 9} (2023) 27}
  [\href{https://arxiv.org/abs/2208.13712}{{\ttfamily 2208.13712}}].

\bibitem{Dixit:2020ymh}
A.V.~Dixit, S.~Chakram, K.~He, A.~Agrawal, R.K.~Naik, D.I.~Schuster et~al.,
  \emph{{Searching for Dark Matter with a Superconducting Qubit}},
  \href{https://doi.org/10.1103/PhysRevLett.126.141302}{\emph{Phys. Rev. Lett.}
  {\bfseries 126} (2021) 141302}
  [\href{https://arxiv.org/abs/2008.12231}{{\ttfamily 2008.12231}}].

\bibitem{PhysRevA.88.041802}
R.~Demkowicz-Dobrza\ifmmode~\acute{n}\else \'{n}\fi{}ski, K.~Banaszek and
  R.~Schnabel, \emph{Fundamental quantum interferometry bound for the
  squeezed-light-enhanced gravitational wave detector geo 600},
  \href{https://doi.org/10.1103/PhysRevA.88.041802}{\emph{Phys. Rev. A}
  {\bfseries 88} (2013) 041802}.

\bibitem{Affeldt:2014rza}
C.~Affeldt et~al., \emph{{Advanced techniques in GEO 600}},
  \href{https://doi.org/10.1088/0264-9381/31/22/224002}{\emph{Class. Quant.
  Grav.} {\bfseries 31} (2014) 224002}.

\bibitem{PhysRevX.13.041021}
{\scshape LIGO O4 Detector Collaboration} collaboration, \emph{Broadband
  quantum enhancement of the ligo detectors with frequency-dependent
  squeezing}, \href{https://doi.org/10.1103/PhysRevX.13.041021}{\emph{Phys.
  Rev. X} {\bfseries 13} (2023) 041021}.

\bibitem{membersoftheLIGOScientific:2024elc}
{\scshape members of the LIGO Scientific\textdagger{}} collaboration,
  \emph{{Squeezing the quantum noise of a gravitational-wave detector below the
  standard quantum limit}},
  \href{https://doi.org/10.1126/science.ado8069}{\emph{Science} {\bfseries 385}
  (2024) ado8069} [\href{https://arxiv.org/abs/2404.14569}{{\ttfamily
  2404.14569}}].

\bibitem{caves1981quantum}
C.M.~Caves, \emph{Quantum-mechanical noise in an interferometer},
  {\emph{Physical Review D} {\bfseries 23} (1981) 1693}.

\bibitem{Kontos}
A.~Thery, \emph{Quantum sensing of axion dark matter with a phase resolved
  haloscope},
  {\emph{https://indico.in2p3.fr/event/32058/contributions/139700/attachments/86008/129188/talk\_Thery\_ENSb.pdf}
  }.

\bibitem{Colpi:2024xhw}
M.~Colpi et~al., \emph{{LISA Definition Study Report}},
  \href{https://arxiv.org/abs/2402.07571}{{\ttfamily 2402.07571}}.

\bibitem{Sesana:2019vho}
A.~Sesana et~al., \emph{{Unveiling the gravitational universe at $\mu$-Hz
  frequencies}}, \href{https://doi.org/10.1007/s10686-021-09709-9}{\emph{Exper.
  Astron.} {\bfseries 51} (2021) 1333}
  [\href{https://arxiv.org/abs/1908.11391}{{\ttfamily 1908.11391}}].

\bibitem{Ruan:2020smc}
W.-H.~Ruan, C.~Liu, Z.-K.~Guo, Y.-L.~Wu and R.-G.~Cai, \emph{{The LISA-Taiji
  network}}, \href{https://doi.org/10.1038/s41550-019-1008-4}{\emph{Nature
  Astron.} {\bfseries 4} (2020) 108}
  [\href{https://arxiv.org/abs/2002.03603}{{\ttfamily 2002.03603}}].

\bibitem{Baker:2019pnp}
J.~Baker et~al., \emph{{Space Based Gravitational Wave Astronomy Beyond LISA}},
  {\emph{Bull. Am. Astron. Soc.} {\bfseries 51} (2019) 243}
  [\href{https://arxiv.org/abs/1907.11305}{{\ttfamily 1907.11305}}].

\bibitem{Kawamura:2011zz}
S.~Kawamura et~al., \emph{{The Japanese space gravitational wave antenna:
  DECIGO}}, \href{https://doi.org/10.1088/0264-9381/28/9/094011}{\emph{Class.
  Quant. Grav.} {\bfseries 28} (2011) 094011}.

\bibitem{TianQin:2020hid}
{\scshape TianQin} collaboration, \emph{{The TianQin project: current progress
  on science and technology}},
  \href{https://doi.org/10.1093/ptep/ptaa114}{\emph{PTEP} {\bfseries 2021}
  (2021) 05A107} [\href{https://arxiv.org/abs/2008.10332}{{\ttfamily
  2008.10332}}].

\bibitem{Kuns:2019upi}
K.A.~Kuns, H.~Yu, Y.~Chen and R.X.~Adhikari, \emph{{Astrophysics and cosmology
  with a decihertz gravitational-wave detector: TianGO}},
  \href{https://doi.org/10.1103/PhysRevD.102.043001}{\emph{Phys. Rev. D}
  {\bfseries 102} (2020) 043001}
  [\href{https://arxiv.org/abs/1908.06004}{{\ttfamily 1908.06004}}].

\bibitem{Reitze:2019iox}
D.~Reitze et~al., \emph{{Cosmic Explorer: The U.S. Contribution to
  Gravitational-Wave Astronomy beyond LIGO}}, {\emph{Bull. Am. Astron. Soc.}
  {\bfseries 51} (2019) 035}
  [\href{https://arxiv.org/abs/1907.04833}{{\ttfamily 1907.04833}}].

\bibitem{Pitkin:2011yk}
M.~Pitkin, S.~Reid, S.~Rowan and J.~Hough, \emph{{Gravitational Wave Detection
  by Interferometry (Ground and Space)}},
  \href{https://doi.org/10.12942/lrr-2011-5}{\emph{Living Rev. Rel.} {\bfseries
  14} (2011) 5} [\href{https://arxiv.org/abs/1102.3355}{{\ttfamily
  1102.3355}}].

\bibitem{Abbott_2019}
B.~Abbott, R.~Abbott, T.~Abbott, S.~Abraham, F.~Acernese, K.~Ackley et~al.,
  \emph{Search for the isotropic stochastic background using data from advanced
  ligo's second observing run},
  \href{https://doi.org/10.1103/physrevd.100.061101}{\emph{Physical Review D}
  {\bfseries 100} (2019) }.

\bibitem{Holometer:2016qoh}
{\scshape Holometer} collaboration, \emph{{MHz Gravitational Wave Constraints
  with Decameter Michelson Interferometers}},
  \href{https://doi.org/10.1103/PhysRevD.95.063002}{\emph{Phys. Rev. D}
  {\bfseries 95} (2017) 063002}
  [\href{https://arxiv.org/abs/1611.05560}{{\ttfamily 1611.05560}}].

\bibitem{Cerdonio_1997}
M.~Cerdonio, M.~Bonaldi, D.~Carlesso, E.~Cavallini, S.~Caruso, A.~Colombo
  et~al., \emph{The ultracryogenic gravitational-wave detector auriga},
  \href{https://doi.org/10.1088/0264-9381/14/6/016}{\emph{Classical and Quantum
  Gravity} {\bfseries 14} (1997) 1491}.

\bibitem{PhysRevLett.18.498}
J.~Weber, \emph{Gravitational radiation},
  \href{https://doi.org/10.1103/PhysRevLett.18.498}{\emph{Phys. Rev. Lett.}
  {\bfseries 18} (1967) 498}.

\bibitem{PhysRevLett.20.1307}
J.~Weber, \emph{Gravitational-wave-detector events},
  \href{https://doi.org/10.1103/PhysRevLett.20.1307}{\emph{Phys. Rev. Lett.}
  {\bfseries 20} (1968) 1307}.

\bibitem{PhysRevD.68.022001}
{\scshape International Gravitational Event Collaboration} collaboration,
  \emph{Methods and results of the igec search for burst gravitational waves in
  the years 1997--2000},
  \href{https://doi.org/10.1103/PhysRevD.68.022001}{\emph{Phys. Rev. D}
  {\bfseries 68} (2003) 022001}.

\bibitem{PhysRevD.76.102001}
{\scshape IGEC-2 Collaboration} collaboration, \emph{Results of the igec-2
  search for gravitational wave bursts during 2005},
  \href{https://doi.org/10.1103/PhysRevD.76.102001}{\emph{Phys. Rev. D}
  {\bfseries 76} (2007) 102001}.

\bibitem{PhysRevLett.94.241101}
L.~Baggio, M.~Bignotto, M.~Bonaldi, M.~Cerdonio, L.~Conti, P.~Falferi et~al.,
  \emph{3-mode detection for widening the bandwidth of resonant gravitational
  wave detectors},
  \href{https://doi.org/10.1103/PhysRevLett.94.241101}{\emph{Phys. Rev. Lett.}
  {\bfseries 94} (2005) 241101}.

\bibitem{Branca:2016rez}
A.~Branca et~al., \emph{{Search for an Ultralight Scalar Dark Matter Candidate
  with the AURIGA Detector}},
  \href{https://doi.org/10.1103/PhysRevLett.118.021302}{\emph{Phys. Rev. Lett.}
  {\bfseries 118} (2017) 021302}
  [\href{https://arxiv.org/abs/1607.07327}{{\ttfamily 1607.07327}}].

\bibitem{Bonaldi:1998gcg}
M.~Bonaldi, P.~Falferi, R.~Dolesi, M.~Cerdonio and S.~Vitale, \emph{{High Q
  tunable LC resonator operating at cryogenic temperature}},
  \href{https://doi.org/10.1063/1.1149166}{\emph{Rev. Sci. Instrum.} {\bfseries
  69} (1998) 3690}.

\bibitem{Anetsberger_2009}
G.~Anetsberger, O.~Arcizet, Q.P.~Unterreithmeier, R.~Rivi{\`e}re,
  A.~Schliesser, E.M.~Weig et~al., \emph{Near-field cavity optomechanics with
  nanomechanical oscillators},
  \href{https://doi.org/10.1038/nphys1425}{\emph{Nature Physics} {\bfseries 5}
  (2009) 909}.

\bibitem{Kawasaki:1999na}
M.~Kawasaki, K.~Kohri and N.~Sugiyama, \emph{{Cosmological constraints on late
  time entropy production}},
  \href{https://doi.org/10.1103/PhysRevLett.82.4168}{\emph{Phys. Rev. Lett.}
  {\bfseries 82} (1999) 4168}
  [\href{https://arxiv.org/abs/astro-ph/9811437}{{\ttfamily
  astro-ph/9811437}}].

\bibitem{Kawasaki:2000en}
M.~Kawasaki, K.~Kohri and N.~Sugiyama, \emph{{MeV scale reheating temperature
  and thermalization of neutrino background}},
  \href{https://doi.org/10.1103/PhysRevD.62.023506}{\emph{Phys. Rev. D}
  {\bfseries 62} (2000) 023506}
  [\href{https://arxiv.org/abs/astro-ph/0002127}{{\ttfamily
  astro-ph/0002127}}].

\bibitem{Hannestad:2004px}
S.~Hannestad, \emph{{What is the lowest possible reheating temperature?}},
  \href{https://doi.org/10.1103/PhysRevD.70.043506}{\emph{Phys. Rev. D}
  {\bfseries 70} (2004) 043506}
  [\href{https://arxiv.org/abs/astro-ph/0403291}{{\ttfamily
  astro-ph/0403291}}].

\bibitem{Planck:2018vyg}
{\scshape Planck} collaboration, \emph{{Planck 2018 results. VI. Cosmological
  parameters}},
  \href{https://doi.org/10.1051/0004-6361/201833910}{\emph{Astron. Astrophys.}
  {\bfseries 641} (2020) A6}
  [\href{https://arxiv.org/abs/1807.06209}{{\ttfamily 1807.06209}}].

\bibitem{Servant:2023tua}
G.~Servant and P.~Simakachorn, \emph{{Ultrahigh frequency primordial
  gravitational waves beyond the kHz: The case of cosmic strings}},
  \href{https://doi.org/10.1103/PhysRevD.109.103538}{\emph{Phys. Rev. D}
  {\bfseries 109} (2024) 103538}
  [\href{https://arxiv.org/abs/2312.09281}{{\ttfamily 2312.09281}}].

\bibitem{Vinante:2006uk}
{\scshape AURIGA} collaboration, \emph{{Present performance and future upgrades
  of the AURIGA capacitive readout}},
  \href{https://doi.org/10.1088/0264-9381/23/8/S14}{\emph{Class. Quant. Grav.}
  {\bfseries 23} (2006) S103}.

\bibitem{Goryachev:2021zzn}
M.~Goryachev, W.M.~Campbell, I.S.~Heng, S.~Galliou, E.N.~Ivanov and M.E.~Tobar,
  \emph{{Rare Events Detected with a Bulk Acoustic Wave High Frequency
  Gravitational Wave Antenna}},
  \href{https://doi.org/10.1103/PhysRevLett.127.071102}{\emph{Phys. Rev. Lett.}
  {\bfseries 127} (2021) 071102}
  [\href{https://arxiv.org/abs/2102.05859}{{\ttfamily 2102.05859}}].

\bibitem{DMRadio:2022jfv}
{\scshape DMRadio} collaboration, \emph{{Proposal for a definitive search for
  GUT-scale QCD axions}},
  \href{https://doi.org/10.1103/PhysRevD.106.112003}{\emph{Phys. Rev. D}
  {\bfseries 106} (2022) 112003}
  [\href{https://arxiv.org/abs/2203.11246}{{\ttfamily 2203.11246}}].

\bibitem{MARGOLUS1998188}
N.~Margolus and L.B.~Levitin, \emph{The maximum speed of dynamical evolution},
  \href{https://doi.org/https://doi.org/10.1016/S0167-2789(98)00054-2}{\emph{Physica
  D: Nonlinear Phenomena} {\bfseries 120} (1998) 188}.

\bibitem{Willke_2002}
B.~Willke, P.~Aufmuth, C.~Aulbert, S.~Babak, R.~Balasubramanian, B.W.~Barr
  et~al., \emph{The geo 600 gravitational wave detector},
  \href{https://doi.org/10.1088/0264-9381/19/7/321}{\emph{Classical and Quantum
  Gravity} {\bfseries 19} (2002) 1377}.

\bibitem{PhysRevLett.101.101101}
T.~Akutsu, S.~Kawamura, A.~Nishizawa, K.~Arai, K.~Yamamoto, D.~Tatsumi et~al.,
  \emph{Search for a stochastic background of 100-mhz gravitational waves with
  laser interferometers},
  \href{https://doi.org/10.1103/PhysRevLett.101.101101}{\emph{Phys. Rev. Lett.}
  {\bfseries 101} (2008) 101101}.

\bibitem{Schmieden:2023fzn}
K.~Schmieden and M.~Schott, \emph{{A Global Network of Cavities to Search for
  Gravitational Waves (GravNet): A novel scheme to hunt gravitational waves
  signatures from the early universe}},
  \href{https://doi.org/10.22323/1.449.0102}{\emph{PoS} {\bfseries EPS-HEP2023}
  (2024) 102} [\href{https://arxiv.org/abs/2308.11497}{{\ttfamily
  2308.11497}}].

\bibitem{Navarro:2023eii}
P.~Navarro, B.~Gimeno, J.~Monz\'o-Cabrera, A.~D\'\i{}az-Morcillo and D.~Blas,
  \emph{{Study of a cubic cavity resonator for gravitational waves detection in
  the microwave frequency range}},
  \href{https://doi.org/10.1103/PhysRevD.109.104048}{\emph{Phys. Rev. D}
  {\bfseries 109} (2024) 104048}
  [\href{https://arxiv.org/abs/2312.02270}{{\ttfamily 2312.02270}}].

\bibitem{Maggiore:1999vm}
M.~Maggiore, \emph{{Gravitational wave experiments and early universe
  cosmology}}, \href{https://doi.org/10.1016/S0370-1573(99)00102-7}{\emph{Phys.
  Rept.} {\bfseries 331} (2000) 283}
  [\href{https://arxiv.org/abs/gr-qc/9909001}{{\ttfamily gr-qc/9909001}}].

\bibitem{Cowan:2010js}
G.~Cowan, K.~Cranmer, E.~Gross and O.~Vitells, \emph{{Asymptotic formulae for
  likelihood-based tests of new physics}},
  \href{https://doi.org/10.1140/epjc/s10052-011-1554-0}{\emph{Eur. Phys. J. C}
  {\bfseries 71} (2011) 1554}
  [\href{https://arxiv.org/abs/1007.1727}{{\ttfamily 1007.1727}}].

\bibitem{d543aecb-cd73-36d5-9101-f08a74f8e8c6}
S.S.~Wilks, \emph{The large-sample distribution of the likelihood ratio for
  testing composite hypotheses}, {\emph{The Annals of Mathematical Statistics}
  {\bfseries 9} (1938) 60}.

\bibitem{wald1943tests}
A.~Wald, \emph{Tests of Statistical Hypotheses Concerning Several Parameters
  when the Number of Observations is Large}, American Mathematical Society
  (1943).

\bibitem{Marzlin:1994ia}
K.-P.~Marzlin, \emph{{Fermi coordinates for weak gravitational fields}},
  \href{https://doi.org/10.1103/PhysRevD.50.888}{\emph{Phys. Rev. D} {\bfseries
  50} (1994) 888} [\href{https://arxiv.org/abs/gr-qc/9403044}{{\ttfamily
  gr-qc/9403044}}].

\end{thebibliography}\endgroup
\bibliographystyle{jhep}
\end{document}